\documentclass[12pt,subeqn,a4paper]{article}

\usepackage{amssymb,amsmath,amsfonts,amsthm, amscd, mathrsfs,helvet}
\usepackage{bm}
\usepackage{graphicx,verbatim}
\usepackage{psfrag}
\usepackage[all]{xy}

\def\r2{\sqrt{2}}

\def\bea{\begin{eqnarray} }
\def\eea{\end{eqnarray}}
\def\be{\begin{equation} }
\def\ee{\end{equation}}
\def\nn{\nonumber}

\begin{document}
\newcommand{\nd}[1]{/\hspace{-0.5em} #1}
\begin{titlepage}
\begin{flushright}
{\bf August 2010} \\ 
%DAMTP-\\
%SWAT-\\ 
%hep-th/yymmnnn \\
\end{flushright}
\begin{centering}
\vspace{.2in}
 {\large {\bf Spiky Strings and Giant Holes}}

\vspace{.3in}

Nick Dorey and Manuel Losi\\
\vspace{.1 in}
DAMTP, Centre for Mathematical Sciences \\ 
University of Cambridge, Wilberforce Road \\ 
Cambridge CB3 0WA, UK \\
\vspace{.2in}
%and \\ 
%\vspace{.2in}
%
%
\vspace{.4in}
{\bf Abstract} \\
\end{centering}

We analyse semiclassical strings in $AdS$ in the limit
of one large spin. 
In this limit, classical string dynamics is
described by a finite number of collective coordinates corresponding to
spikes or cusps of the string. The semiclassical spectrum 
consists of two branches of excitations 
corresponding to ``large'' and ``small'' spikes respectively. We
propose that these states are dual to the excitations known as large and
small holes in the spin chain description of ${\cal N}=4$ SUSY Yang-Mills.   
The dynamics of large spikes in classical string theory 
can be mapped to that of a classical spin
chain of fixed length. In turn, small spikes correspond to classical solitons 
propagating on the background formed by the large spikes. 
We derive the dispersion relation for these excitations directly in
the finite gap formalism.

%\vspace{.05in}
%\baselineskip=.3in
\end{titlepage}
\paragraph{}
\section{Introduction}
\begin{figure}
\centering
\psfrag{S1}{\footnotesize{$\Sigma$}}
\psfrag{S2}{\footnotesize{$\tilde{\Sigma}_{2}$}}
\psfrag{B}{\footnotesize{$\hat{\mathcal{B}}$}} 
\psfrag{A2}{\footnotesize{$\hat{\mathcal{A}}_{2}$}}
\psfrag{Ip}{\footnotesize{$\infty^{+}$}}
\psfrag{Im}{\footnotesize{$\infty^{-}$}}
\includegraphics[width=100mm]{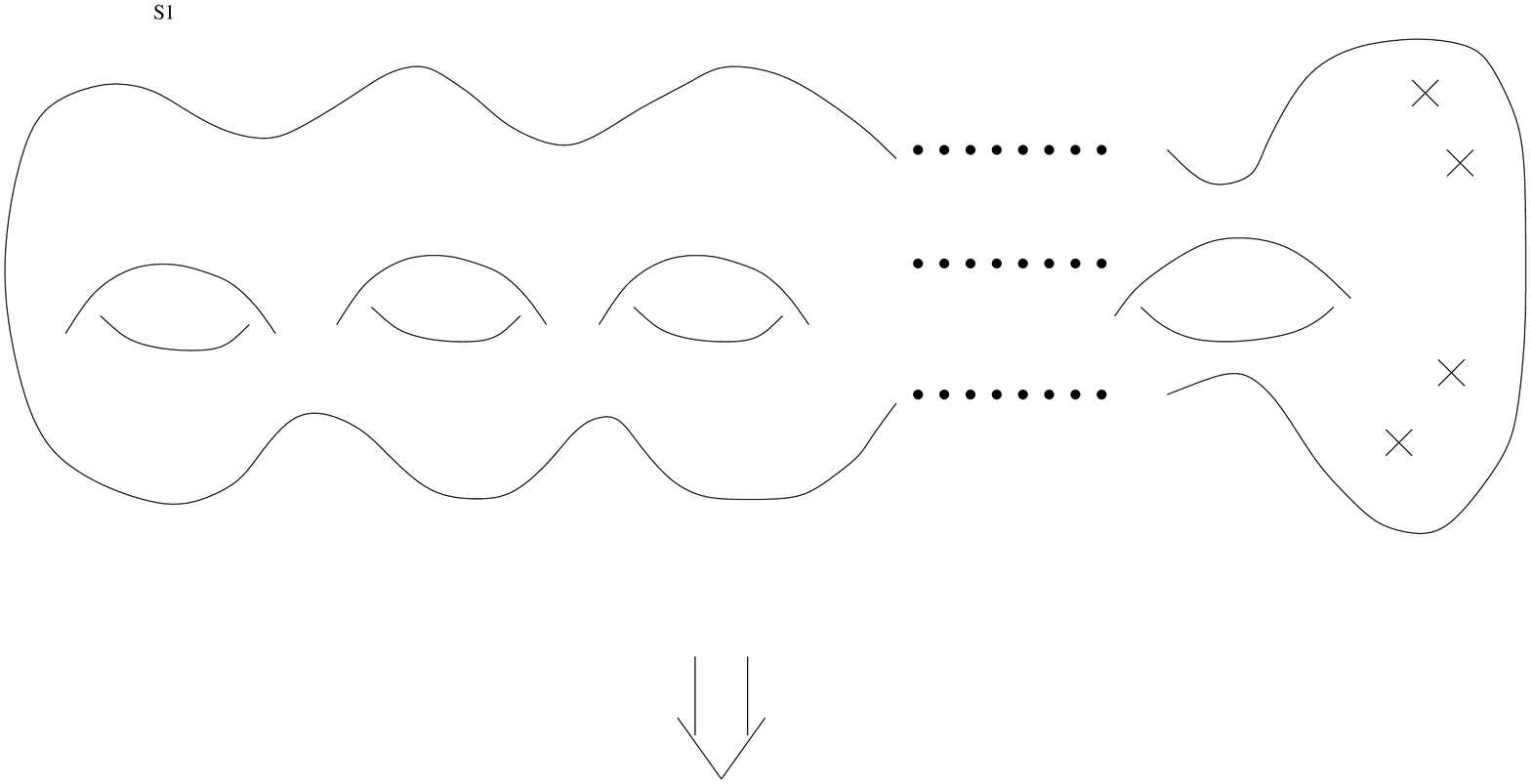}
%\caption{The ``extra'' cycle $\mathcal{A}^{+}_{K/2}$ becomes the sum
%  of the chains $\hat{\mathcal{A}}_{1}$ and $\hat{\mathcal{A}}_{2}$.}
%\label{Gfig1a}
%\end{figure}
%\begin{figure}
\centering
\psfrag{B}{\footnotesize{${\mathcal{C}}$}} 
\psfrag{S1}{\footnotesize{$\tilde{\Sigma}_{1}$}}
\psfrag{S2}{\footnotesize{$\tilde{\Sigma}_{2}$}}
\psfrag{CI}{\footnotesize{$C^{\pm}_{I}$}}
\psfrag{Ip}{\footnotesize{$\infty^{+}$}}
\psfrag{Im}{\footnotesize{$\infty^{-}$}}
\includegraphics[width=100mm]{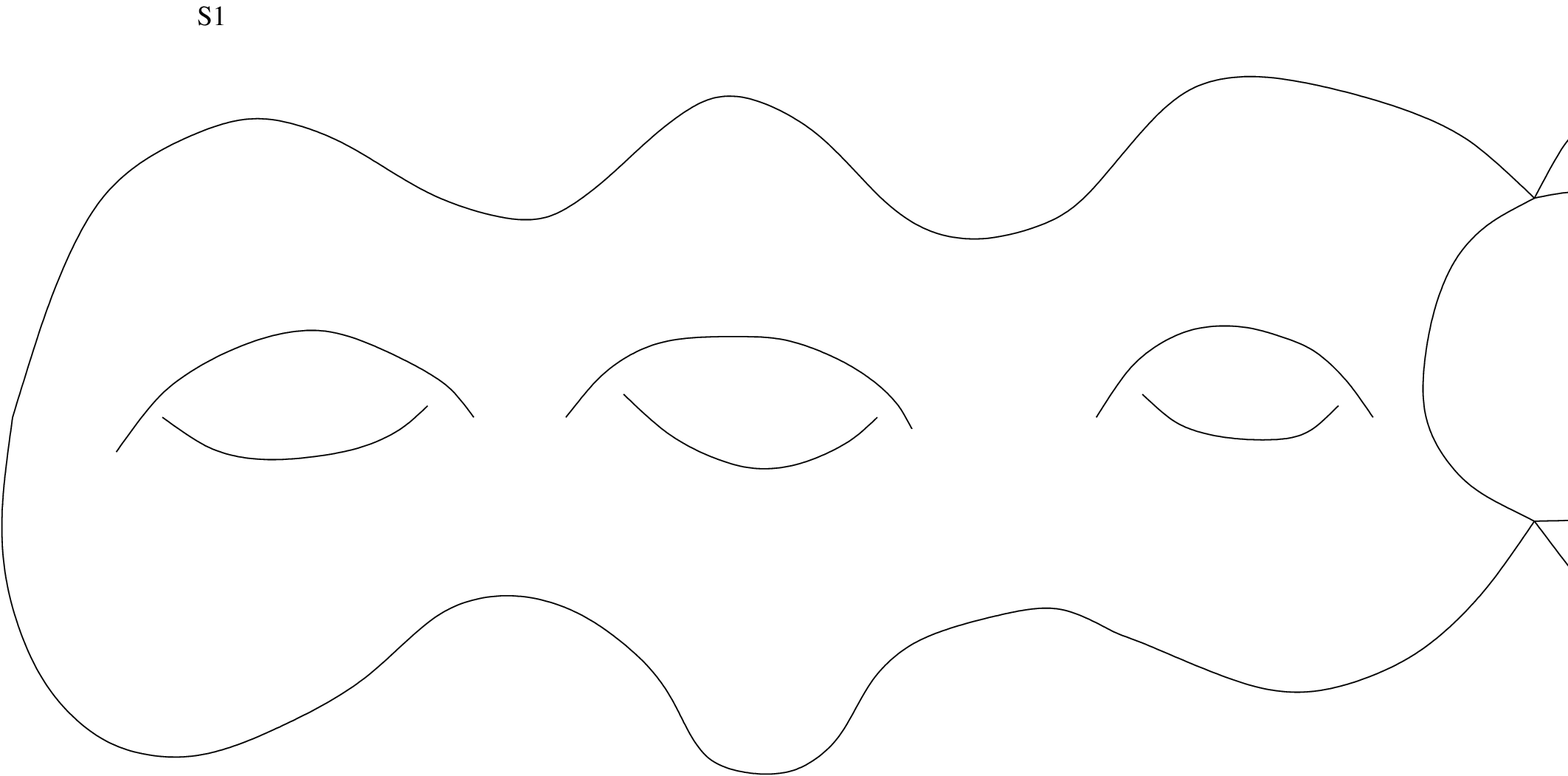}
\caption{The degeneration $\Sigma \rightarrow \tilde{\Sigma}_{1}\cup
\tilde{\Sigma}_{2}$ in the case $K=9, M=4$. As explained in Section 3, crosses denote second order
poles in the differential $dp$ while the blue dots denote simple poles.}
\label{Gfig1b}
\end{figure}

\paragraph{}
The emergent integrability of planar ${\cal N}=4$ SUSY Yang-Mills \cite{MZ,B}
and the dual string theory on $AdS_{5}\times S^{5}$ \cite{MSW,BPR} 
has provided detailed information about the spectra of both theories 
(for a recent review and additional references see 
\cite{Serban}). In both gauge theory and string theory, the problem reduces 
to that of solving an integrable system in $(1+1)$-dimensions. As usual 
for such systems, the spectrum has a simple description for large spatial 
volume in terms of asymptotic states which undergo factorised scattering. 
Much recent progress is based on our knowledge of the 
exact spectrum of excitations \cite{BDS,BS,B1} around the chiral primary
operator\footnote{Here and in the following $Z$ is one of the three 
complex adjoint scalar fields of the ${\cal N}=4$ theory.} 
${\rm Tr}[Z^{J}]$ in the $J\rightarrow\infty$ limit where the 
operator and the dual string become long.  In the spin chain 
description of planar gauge theory, the chiral primary operator 
corresponds to the ferromagnetic groundstate and the relevant excitations 
are magnons. In string theory the dual excitations, known as 
Giant Magnons \cite{HM}, are classical solitons in the worldsheet 
$\sigma$-model. 
\paragraph{}
Recently it has become clear that a description of the
spectrum in terms of asymptotic states and factorised scattering
should also arise in another quite different limit: the limit of large
Lorentz spin. This limit arises for operators containing a large
number of covariant derivatives of the same kind. 
A convenient example is provided by the operators of the $sl(2)$
sector which take the form,    
\bea 
\hat{O} & \sim & {\rm Tr}_{N}\left[\,\mathcal{D}_{+}^{s_{1}}Z\,
\mathcal{D}_{+}^{s_{2}}Z\,\ldots\, \mathcal{D}_{+}^{s_{J}}Z\,\right] 
\label{sl2} 
\eea 
where $\mathcal{D}_{+}$ is the gauge-theory covariant derivative 
in a light-like direction and the total Lorentz spin is 
$S = \sum_{l=1}^{J}\,\,s_{l}$. 
The $S\rightarrow\infty$ limit of the operator spectrum, where the
number of derivatives becomes large, is relevant for the study of Wilson
loops. In particular, the dimension of the large-$S$ groundstate 
is directly related to
the cusp anomalous dimension of a Wilson loop. Moreover 
the spectrum of excitations above this groundstate plays an 
important role in recent work on the Operator
Product Expansion for light-like Wilson loops \cite{AGV}. Intuitively
it is natural to expect that the excitations correspond to 
insertions of $Z$ and other impurities propagating in a ``sea'' of 
derivatives. 
As we review below, this intuition can be made precise in the spin chain
description of the planar gauge theory. In this context, the 
derivatives correspond to magnons which fill a Dirac sea and 
excitations around this groundstate 
are identified as holes in the rapidity distribution. 
\paragraph{}
In this paper we focus on the corresponding states in 
semiclassical string theory. We present evidence that, as in the
case of magnons, the dual states are solitonic excitations 
corresponding to spikes or cusps which propagate on a long string. 
In analogy with the Giant Magnons of \cite{HM}, these states were referred to
as Giant Holes in \cite{GH}. We will argue that the
spectrum contains two branches corresponding to ``large'' and
``small'' spikes respectively\footnote{This terminology is 
potentially confusing. Here the terms 
``large'' and ``small'' refer to $O(S)$ and $O(1)$
 contributions to the string energy in the $S\rightarrow \infty$ limit
 respectively. In contrast the name Giant Hole (like the name Giant
 Magnon) refers to a contribution to the string energy which grows
 like $\sqrt{\lambda}$ in the semiclassical limit 
$\lambda=g^{2}_{\rm YM}N\rightarrow \infty$. As we will see below
Giant Holes can be either ``large'' or ``small''.} 
which separate in the $S\rightarrow \infty$ limit.  
As we review below, 
this precisely mirrors a corresponding distinction between 
``large'' and ``small'' holes in
the gauge theory spin chain. In the following we will present a
detailed analysis of the Hamiltonian dynamics of these states using
the finite gap construction of $AdS$ string solutions. In particular
we give a rigorous derivation of the dispersion relation for
Giant Holes presented in \cite{GH}. 
We will now outline the main ideas and results
beginning with a brief review of one-loop gauge theory in the
large-$S$ limit.           
\paragraph{}
For definiteness we will focus on the $sl(2)$ sector of the 
planar gauge theory, consisting of operators of the form (\ref{sl2})
introduced above. This 
set of operators naturally corresponds to the Hilbert space of a
non-compact spin chain of 
length $J$ with a discrete lowest-weight representation of 
$SL(2,\mathbb{R})$ at each site. 
The one-loop planar dilatation operator in this sector 
coincides with the nearest-neighbour Heisenberg Hamiltonian of the chain. 
The exact spectrum of one-loop anomalous dimensions is determined by 
the Bethe Anzatz which diagonalises the Heisenberg Hamiltonian.    
With this identification, the 
chiral primary operator ${\rm Tr}[Z^{J}]$ corresponds to the 
ferromagnetic groundstate. 
Each derivative corresponds to a magnon 
%The one-loop anomalous dimensions of these operators are
%determined by the energy levels of an integrable ${\rm SL}(2,R)$ 
%spin chain which are determined exactly by the Bethe ansatz. To
%formulate the Bethe ansatz one starts from the ferromagnetic vacuum of
%the spin chain which corresponds to the chiral primary $\hat{O}_{\rm
%  BPS}$. Excitations around this groundstate correspond to magnons
carrying spin $S=1$, as well as conserved energy and momentum 
parametrised in terms of
a complex rapidity variable $u$ as,    
\bea 
p(u)=\frac{1}{i}\log \left(\frac{u-i/2}{u+i/2}\right),  & \qquad{}
\qquad{} & \epsilon(u)= \frac {1}{u^{2}+1/4} \nn \eea
%Each magnon carries spin $S=1$ and corresponds to the insertion of a
%single covariant derivative $\mathcal{D}_{+}$ into the chiral primary
%$\hat{O}_{\rm BPS}$. 
As the interactions of the spin chain are short-range, the total
energy of an $S$-magnon state is, 
\bea E & = & \sum_{l=1}^{S}\,\epsilon(u_{l}) \nn \eea
the rapidities $\{u_{l}\}$ of individual magnons, also known as Bethe
roots, are determined by the
Bethe Ansatz Equations (BAE) which have the form, 
\bea
p(u_{l})J & = & n_{l}\pi \,\,+\,\, \sum_{l'\neq l}^{S}\,
\delta(u_{l}-u_{l'})
\nn \eea 
where $n_{l}$ is an integer known as the {\em mode number} and 
$\delta(u)=i\log[(u+i)/(u-i)]$ is the phase shift for two-magnon
scattering. A state is uniquely
characterised by the corresponding distribution of mode
numbers. The non-compact spin chain in question has some remarkable
properties (for a review see \cite{BGK,FS}). 
First, all Bethe roots are real and no two roots can have the same mode
number. Second, with appropriate conventions, there are always exactly
$J$ holes in the distribution of mode numbers. The holes are
in one-to-one correspondence with the zeros $u=u_{j}$, $j=1,\ldots,J$
of the spin chain transfer matrix. 
\paragraph{}
In the $S\rightarrow \infty$ limit, the number of magnons becomes
large but the number of holes $J$ remains fixed. Hence it is
convenient to reformulate the Bethe ansatz in terms of the holes and
their associated rapidities $\{u_{j}\}$, $j=1,\ldots,J$. Remarkably,
the holes acquire the attributes of particles in this
reformulation. Specifically each hole carries conserved energy and
momentum parametrised in terms of the corresponding rapidity $u$ as,    
\bea 
p(u)=\frac{1}{i}\log \left[\frac{\Gamma(1/2 - iu)}{\Gamma(1/2
    +iu)}\right],  
& \qquad{} & \epsilon(u)= \psi(1/2
    +iu) +\psi(1/2- iu)-2\psi(1) \label{disp1} \eea 
where $\psi(x)=d(\log\Gamma(x))/dx$. Further, the total energy of the
state is essentially the sum of the energies of the individual
holes, 
\bea E & = & 2\log 2 \,\,+\,\,\sum_{j=1}^{J}\, \epsilon(u_{j}) \nn
\eea
Finally the rapidities of the holes are detemined in the $S\rightarrow
\infty$ limit by a dual set of Bethe equations whose explicit form
will not be needed here. Like the original Bethe Ansatz Equations
given above, these equations can be interpreted as the periodicity
conditions for the wavefunction of a multi-particle state where the
individual particles undergo factorised scattering. 
\paragraph{}
The solution of the dual Bethe equations in the $S\rightarrow \infty$
limit was discussed in detail in \cite{BGK}. In this limit, the
excitations are
naturally divided up in ``large'' holes for which $u_{j}\sim S$ and
``small'' holes whose rapidity remains fixed (or goes to zero) as
$S\rightarrow \infty$. Large holes make a logarithmically growing 
contribution to the energy in this limit 
according to the asymptotics of the dispersion relation (\ref{disp1}):
$\epsilon(u) \simeq \log(u)\,\,+\,\,O(u^{0})$. The number $K$ of large
holes lies in the range $2\leq K \leq J$ and states with $K$ large
holes contribute to the anomalous dimension as, 
\bea 
\gamma\,\,\simeq\,\, \Delta - S & = & \frac{\lambda}{4\pi^{2}} \,\,
K\log \, S \qquad{} + \qquad{} {\rm O}(S^{0}) \label{mform} \eea 
\paragraph{}
The large-$S$ limit can also be understood as a semiclassical limit of
the $SL(2,\mathbb{R})$ spin chain where at least some of the
non-compact spins become highly excited and can be replaced by
classical variables. The dynamics of the large
holes is captured precisely by the resulting classical spin chain. The
number $K\geq 2$ of large holes corresponds to the number of classical
spins. For $K>2$, the semiclassical quantisation of the 
chain yields a band of states, labelled by $K-2$ integer valued
quantum numbers, 
corresponding to different configurations of the large
holes. For each state, the $O(S^{0})$ terms in the anomalous dimension  
(\ref{mform}) are captured by the Bohr-Sommerfeld quantisation
of classical orbits. The remaining excitations, the $J-K$ small holes, 
can be thought of as small
excitations around the background formed by the large holes. In the
groundstate, each small hole makes a contribution of order $1/\log S$ 
to the anomalous dimension as $S\rightarrow \infty$. 
\paragraph{}
We will now analyse the same $S\rightarrow
\infty$ limit in semiclassical string theory on $AdS_{5}\times
S^{5}$. The dual to the gauge theory operator of lowest dimension in
the $sl(2)$ sector is 
provided by a folded string, spinning in an $AdS_{3}$ subspace of
$AdS_{5}$ \cite{GKP}. The $AdS$ angular momentum is identified with the
conformal spin $S$ of the dual gauge theory denoted by the same
letter. To account for the R-charge of the dual gauge theory
operators, the string solution we consider also carries 
$J$ units of angular momentum around
an $S^{1}$ on $S^{5}$.  
As $S\rightarrow\infty$ with $J$ fixed, the endpoints of the string
approach the boundary of $AdS$ and the string energy scales
logarithmically with $S$ in qualitative agreement with gauge theory; 
\bea \gamma & = &  2\Gamma(\lambda)\,\log(S) \qquad{}+ \qquad{} O(S^{0})
\nn \eea  
For large $\lambda$, where the semiclassical approximation is valid,
the coupling-dependent prefactor $\Gamma(\lambda)$ takes the value
$\sqrt{\lambda}/2\pi$. However, it can be calculated exactly \cite{BES}, 
for all values of $\lambda$, using the Asymptotic Bethe ansatz and
correctly interpolates to the corresponding one-loop prefactor 
in (\ref{mform}) (in the ground-state case $K=2$) when $\lambda$ is small. 
Importantly, the quantity $\Gamma(\lambda)$ also coincides with the 
cusp anomalous dimension of a light-like Wilson loop. 
\paragraph{}
Having identified the folded spinning string as the state of lowest
energy at fixed large $S$, our main goal is to understand the spectrum of 
excitations over this groundstate. The spectrum of small 
fluctuations around the GKP string is well understood \cite{AM2}, 
however we will
focus on classical excitations of the string which naturally carry energy of
order $\sqrt{\lambda}$ in the limit $\sqrt{\lambda}\rightarrow
\infty$. The key objects we will study are the spikes or
cusps\footnote{More precisely, these are
points on the string at which, in an appropriate gauge, the spacelike
worldsheet derivatives of the spacetime coordinates of the string
vanish.} of the semiclassical string in $AdS$.  
Like the holes described 
above, the spikes are naturally divided into ``large'' and
``small''. Large spikes are those which approach the
boundary of $AdS$ as $S\rightarrow\infty$. Like the corresponding
holes, they contribute an amount of order $\log S$ to the string
energy. From this point of view, the GKP folded string has two large
spikes. Spinning string solutions with 
$K>2$ large spikes located at the vertices of a regular polygon were
given explicitly by Kruczenski \cite{Kruc}. A detailed correspondence
between the dynamics of large spikes and that of the classical spin
chain describing the large holes of the gauge theory spin chain was
given in \cite{DS,DL1} and will be discussed further below.
\paragraph{}
The excitation spectrum of the folded string also contains ``small''
spikes which contribute to the string energy at order $S^{0}$ in the
large-$S$ limit. These
states correspond to classical solitons which propagate on the GKP
folded string (or, more generally, on a solution with any number of
large spikes). Two spike solutions of this kind were 
first constructed in \cite{Jev1} using Pohlmeyer reduction 
\cite{Pohlmeyer,otherPohl}. Generic multi-spike configurations
were constructed using the inverse scattering method in 
\cite{Jev2}. The dispersion relation for these excitations was 
recently derived in \cite{GH}. In appropriate worldsheet coordinates, the 
conserved energy and momentum are given in terms of the soliton
velocity as, 
 \bea E_{\rm sol}(v) & = & \frac{\sqrt{\lambda}}{2\pi}
\left[\frac{1}{2}\log\left(\frac{1+\sqrt{1-v^{2}}}
{1-\sqrt{1-v^{2}}}\right)\,\,-\,\,
\sqrt{1-v^{2}}\right] \nn \\ 
P_{\rm sol}(v) & = &
\frac{\sqrt{\lambda}}{2\pi}\left[\frac{\sqrt{1-v^{2}}}{v}\,\,-\,\,
{\rm Tan}^{-1}\left(\frac{\sqrt{1-v^{2}}}{v}\right)\right] 
\label{disp2} \eea
The dispersion relation interpolates between relativistic behaviour
$E_{\rm sol}\simeq P_{\rm sol}$ at low energy, and logarithmic
behaviour at high energy, $|E_{\rm sol}| \simeq 
(\sqrt{\lambda}/2\pi) \log |P_{\rm sol}|$ 
which is related to the $\log S$ scaling of the string energy
described above. Our main goal here is to study the dynamics of these
spikes using the finite gap construction.  
%The above dispersion relation $E_{\rm sol}(P_{\rm sol})$
%takes the relativistic form at low values of the momentum: $E_{\rm
%  sol} \simeq |P_{\rm sol}| + O(|P_{\rm sol}|^{5/3})$ for $|P_{\rm
%  sol}| \ll 1$. This property was also observed in the case of the
%Giant Magnons studied by Hofman and Maldacena \cite{HM}, which arise
%as solitonic excitations above the BMN vacuum. At the opposite end of
%the momentum scale, when $|P_{\rm sol}| \gg 1$, the dispersion
%relation displays the same leading behaviour $|E_{\rm sol}| \simeq
%(\sqrt{\lambda}/2\pi) \log |P_{\rm sol}| + O(|P_{\rm sol}|^0)$ as we
%saw above for the holes of the ${\rm SL}(2,R)$ spin chain
%\eqref{disp1}. Because of these analogies, the name ``Giant Holes''
%was proposed for these excitations. 
%Finally, the leading order quantisation condition
%\bea
%P_{\rm sol}(v)\cdot 2\log S & \in & 2\pi\,\,\mathbb{Z}
%\label{quant} \eea
%yields the semiclassical spectrum at large $S$.
\paragraph{}
The finite gap formalism provides a construction of generic classical
string solutions on $AdS_{3}\times S^{1}$ \cite{KMMZ,KZ}. 
The construction starts
from the Lax reformulation of the equations of motion as the flatness
condition for a one-parameter family of connections on the
worldsheet. Solutions are characterised by the analytic behaviour of
the Lax connection as a function of the complex spectral
parameter. More precisely, the 
monodromy of the Lax connection around the closed string gives rise to a
differential $dp$, known as the quasi-momentum, which is meromophic on
an auxiliary Riemann surface or spectral curve. 
The classical spectrum of the conserved charges for a particular
solution is encoded in the spectral curve. The space of string
solutions is divided into sectors according to the genus of the
corresponding curve which is equal to $K-1$ where $K$ is the number of 
gaps in the spectrum of the auxiliary linear problem. 
This provides a non-linear analog of the mode
expansion of string solutions in flat space. In particular, solutions
corresponding to a curve of genus $K-1$ are continuously related to
linearised solutions with $K$ active modes (For more details see
\cite{KZ, DV1}). The space of $K$-gap solutions inherits a symplectic
structure from the string $\sigma$-model. This yields a Hamiltonian
description of classical string dynamics which can also be quantised to
describe the leading semiclassical approximation to the string
spectrum for $\lambda >>1$.     
\paragraph{}
In this paper we will analyse the $S\rightarrow \infty$ limit of a
generic string solution in $AdS_{3}\times S^{1}$ in terms of the
corresponding spectral curve. As we review below, the Riemann surface
for a generic $K$-gap solution is a hyperelliptic curve of genus $K-1$
which admits a normalised Abelian differential of the second
kind $dp$ with prescribed behaviour at singular points. The 
existence of such a differential puts constraints on the moduli of a
general hyperelliptic curve, leaving a $K-1$ dimensional moduli space of 
solutions. We start from a generic curve obeying these conditions and
consider its behaviour as $S\rightarrow \infty$. The most
general large-$S$ limit gives rise to a degeneration of the curve
depicted in Figure \ref{Gfig1b}, 
\begin{eqnarray}
\Sigma & \rightarrow &  \tilde{\Sigma}_{1} \cup \tilde{\Sigma}_{2} 
\end{eqnarray}
Here $\tilde{\Sigma}_{1}$ has genus $K-M-2$ for $M=0,1,\ldots,K-2$,
while $\tilde{\Sigma}_{2}$ has genus zero but $M$ additional punctures
(denoted by blue dots in Figure \ref{Gfig1b}) which correspond to 
simple poles in
the quasi-momentum $dp$. We will show in detail that the two
components $\tilde{\Sigma}_{1}$ and $\tilde{\Sigma}_{2}$ describe the
decoupled dynamics of ``large'' and ``small'' spikes respectively. 
\begin{figure}
\centering
\includegraphics[width=100mm]{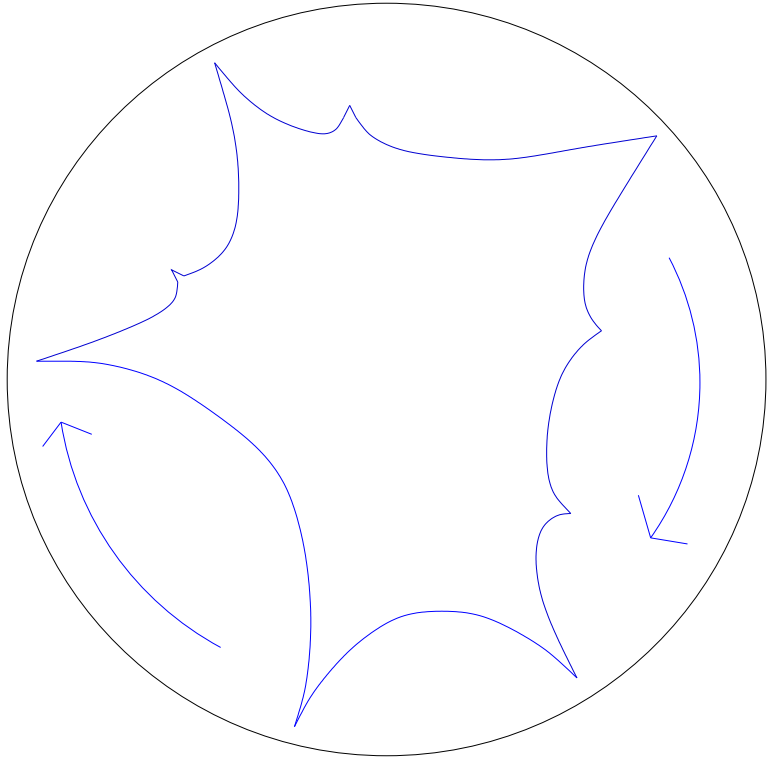}
\caption{Small spikes propagating on a background of big spikes. The
  specific solution illustrated corresponds to the case $K=9$, $M=4$ 
and the curve shown in Figure 1.}
\label{SGMfig7}
\end{figure}
\paragraph{}
The degeneration considered here is a generalisation of the limit
considered in \cite{DS} (which corresponds to the case $M=0$) and our
analysis of the component $\tilde{\Sigma}_{1}$ closely follows this
reference. In particular we find that the period conditions for
$\tilde{\Sigma}_{1}$ can be
solved explicitly and the resulting curve precisely coincides with
that of a classical $SL(2,\mathbb{R})$ Heisenberg spin chain of length
$K-M$. Further, the semiclassical spectrum of string states
corresponding to this sector coincides with the semiclassical spectrum
of the spin chain. The same chain emerges directly in the analysis of
string solutions with $K-M$ large spikes \cite{DL1}. 
As worked out in detail in
\cite{DL1}, the spin degrees of freedom correspond to the localised
worldsheet Noether charges at each cusp of the string. Further, the moduli of
$\tilde{\Sigma}_{1}$ correspond directly to the angular positions of
the large spikes.  
\paragraph{}
The main new ingredient in the present analysis is the new sectors
with $M>0$ where there are $M$ additional simple poles in the
quasi-momentum $dp$ on $\tilde{\Sigma}_{2}$. The poles arise in a
standard way from the collision of branch points in the spectral data
and are naturally associated with solitons propagating on the string.   
We will derive the corresponding semiclassical spectrum of string
states and obtain a detailed match to the properties of the small
spikes described above. In particular, we reproduce the dispersion
relation (\ref{disp2}). The resulting picture of the general large-$S$
state described by the curve $\Sigma$ is of $M$ small spikes
propagating on the background provided by $K-M$ large spikes. A
generic solution of this type is illustrated in Figure \ref{SGMfig7}. 
The degeneration of the curve shown in Figure \ref{Gfig1b} corresponds 
to the decoupling of these two
branches of the spectrum as $S\rightarrow \infty$. 
\paragraph{}
The remainder of the paper is organised as follows. In the next
section we review the basic set-up of the finite gap formalism for
strings on $AdS_{3}\times S^{1}$. In Section 3 we consider the 
$S\rightarrow \infty$ limit of a generic $K$-gap string solution with $AdS$ 
angular momentum $S$, in particular we review the analysis of
\cite{DS} which leads to a precise identification of the dynamics of
large spikes with that of the classical gauge theory spin chain. 
We then generalise this analysis to include configurations
corresponding to small spikes. We derive the dispersion
relation for small spikes directly from the corresponding limit of the
spectral curve. Some details of the calulation are relegated to an 
appendix.

\section{The finite gap construction}
\label{sec:finite-gap}
\paragraph{}
We begin by focusing on classical bosonic string theory
on $AdS_{3}\times S^{1}$. This is a sub-sector of classical 
string theory on the full $AdS_{5}\times S^{5}$ geometry which
contains in particular the string duals to the $sl(2)$ sector
operators of the form (\ref{sl2}). Here the $U(1)$ R-charge $J$ of the
gauge theory corresponds
to momentum in the $S^{1}$ direction and the conformal spin $S$
corresponds to angular momentum in $AdS_{3}$. The string energy
$\Delta$ corresponds to translations of the global time in $AdS_{3}$. 
\paragraph{}
The simplest string solution is the so-called BMN groundstate: 
a pointlike string orbiting $S^{1}$ at the speed of light. This
corresponds to the ferromagnetic groundstate of the dual gauge theory
spin chain and has exact energy $\Delta=J$.  
\paragraph{}
We introduce string
worldsheet coordinates $\sigma \sim \sigma + 2\pi$ and $\tau$ and 
the corresponding lightcone coordinates $\sigma_{\pm}=(\tau\pm
\sigma)/2$ and we define lightcone derivatives
$\partial_{\pm}=\partial_{\tau} \pm \partial_{\sigma}$. The space-time
coordinates correspond to fields on the string worldsheet: we
introduce $\phi(\sigma,\tau)\in S^{1}$ and parametrize $AdS_{3}$ with
a group-valued field $g(\sigma,\tau)\in SU(1,1)\simeq
AdS_{3}$. The $SU(1,1)_{R}\times SU(1,1)_{L}$
isometries of $AdS_{3}$ correspond to left and right group
multiplication. The Noether current corresponding to right
multiplication is $j_{\pm}=g^{-1}\partial_{\pm}g$. Following \cite{KZ}, we
work in static, conformal gauge with a flat worldsheet metric and set, 
\bea \phi(\sigma,\tau) & = & \frac{J}{\sqrt{\lambda}}\,\,\tau \nn \eea 
In this gauge, the string action becomes that of the $SU(1,1)$ 
Principal Chiral model, 
\bea 
S_{\sigma} & = & \frac{\sqrt{\lambda}}{4\pi}\, \int_{0}^{2\pi}
\,d\sigma\, \frac{1}{2}{\rm tr}_{2}\left[j_{+}j_{-}\right] \label{sigma} \eea 
\paragraph{}
As usual we expect a string moving in four spacetime dimensions to
have two transverse degrees of freedom. Expanding around the BMN
groundstate solution described above, we find two oscillation 
modes of equal mass carrying charges $S=\pm 1$. This is the usual
transverse spectrum of the $AdS_{5}\times S^{5}$ string restricted to
a bosonic $AdS_{3}\times S^{1}$ subspace. The two modes are dual to
magnons of the gauge theory spin chain and correspond to insertions of
light-cone covariant derivatives $\mathcal{D}_{+}$ and $\mathcal{D}_{-}$ in the
BMN groundstate operator. The former excitations correspond to the
$sl(2)$ sector operators identified above. The lightcone derivative 
$\mathcal{D}_{-}$ also generates a closed subsector of operators described
by a Heisenberg spin at one loop, but operators containing both types
of excitation do not form a closed subsector.         
\paragraph{}
The equations of motion corresponding to (\ref{sigma}) 
are integrable by virtue of their
equivalence to the consistency conditions for the auxiliary linear problem, 
%\bea 
%\left[\partial_{\sigma} \,\,+ \,\, \left( \frac{j_{+}}{x-1} \,\,+\,\,
%\frac{j_{-}}{x+1}\right )\right]\cdot \vec{\Psi} & = & 0 \nn \\ 
%\left[\partial_{\tau} \,\,+ \,\, \left( \frac{j_{+}}{x-1} \,\,-\,\,
%\frac{j_{-}}{x+1}\right )\right]\cdot \vec{\Psi} & = & 0 \nn \eea
\bea 
\left[\partial_{\sigma} \,\,+ \,\, \frac{1}{2} \left( \frac{j_{+}}{1-x} \,\,-\,\,
\frac{j_{-}}{1+x}\right )\right]\cdot \vec{\Psi} & = & 0 \nn \\ 
\left[\partial_{\tau} \,\,+ \,\, \frac{1}{2} \left( \frac{j_{+}}{1-x} \,\,+\,\,
\frac{j_{-}}{1+x}\right )\right]\cdot \vec{\Psi} & = & 0 \nn \eea
where $x\in \mathbb{C}$ is a complex spectral parameter. String motion 
is also subject to the Virasoro constraint, 
%\bea 
%\frac{1}{2}{\rm tr}_{2}\left[j_{\pm}^{2}\right] & = &
%\frac{J^{2}}{\lambda}
%\nn \eea  
\bea 
\frac{1}{2}{\rm tr}_{2}\left[j_{\pm}^{2}\right] & = & -
\frac{J^{2}}{\lambda}
\nn \eea  
 \paragraph{}
Classical integrability of string theory on $AdS_{3}\times S^{1}$
follows from the construction of the monodromy matrix \cite{BPR}, 
\bea 
\Omega\left[x;\tau \right] & = &
\,\,\mathcal{P}\,\,\exp\left[\frac{1}{2}
\int_{0}^{2\pi}\,d\sigma \,\, \left( \frac{j_{+}}{x-1} \,\,+\,\,
\frac{j_{-}}{x+1}\right )\right] \,\,\,\,\in\,\,\,\, SU(1,1)
\nn \eea 
whose eigenvalues $w_{\pm}=\exp(\pm i\,p(x))$ are
$\tau$-independent for all values of the spectral parameter $x$. It is
convenient to consider the analytic continuation of the monodromy
matrix $\Omega[x;\tau]$ and of the quasi-momentum $p(x)$ to complex
values of $x$. In this case $\Omega$ will take values in
$SL(2,\mathbb{C})$ and appropriate reality conditions must be imposed
to recover the physical case.   
\paragraph{}
The eigenvalues $w_{\pm}(x)$ are two branches of an analytic
function defined on the spectral curve, 
\bea 
\Sigma_{\Omega}\,\,: \qquad{} \qquad{} 
{\rm det}\left(w \mathbb{I} \,\,-\,\,\Omega[x;\tau]
\right)\,\,=\,\, w^{2}\,\,-\,\,2\cos p(x) \, w\,\,+\,\,1  & =
& 0 \qquad{} w, \,\,x\in \mathbb{C} \nn \eea 
This curve corresponds to a double cover of the complex $x$-plane with
branch points at the simple zeros of the discriminant
$D=4\sin^{2}p(x)$. In addition the monodromy matrix defined above is
singular at the points above $x=\pm 1$. Using the Virasoro constraint,
one may show that $p(x)$ has simple poles at these points,  
\bea 
p(x) & \sim & \frac{\pi J}{\sqrt{\lambda}}\frac{1}{(x\pm 1)} \,+\, 
O\left((x\pm 1)^{0}\right) \label{simp} \eea
as $x\rightarrow \mp 1$. Hence
the discriminant $D$ has essential singularities at $x=\pm 1$ and $D$
must therefore have an infinite number of zeros which
accumulate at these points. Formally we may represent the discriminant
as a product over its zeros and write the spectral curve as 
\bea 
\Sigma_{\Omega}\,\,: \qquad{} \qquad{}  {y}^{2}_{\Omega}\,\,=
\,\,4\sin^{2}p(x) & = & \prod_{j=1}^{\infty} \left(x-x_{i}\right) \nn
\eea 
For generic solutions the points $x=x_{i}$ are distinct and the curve 
${\Sigma}_{\Omega}$ has infinite genus. 
\paragraph{}
In order to make progress it is necessary to focus on 
solutions for which the discriminant has only a finite number $2K$ of 
simple zeros and the spectral curve ${\Sigma}_{\Omega}$ has finite genus. 
The infinite number of additional zeros of the discriminant $D$ must
then have multiplicity two or higher. 
These are known as {\rm finite gap solutions}\footnote{Strictly
  speaking these are not generic solutions of the string equations of
  motion. However, as $K$ can be
arbitrarily large, it is reasonable to expect that generic solutions
could be obtained by an appropriate $K\rightarrow \infty$ limit.}.  
In this case, $dp$ is a
meromorphic differential on the hyperelliptic curve 
of genus $g=K-1$ which is obtained by removing the double points of 
$\hat{\Sigma}$ (see \cite{DV1}).

\subsection*{Finite-gap review}
\begin{figure}
\centering
\psfrag{x}{\footnotesize{$x$}}
\psfrag{m1}{\footnotesize{$-1$}}
\psfrag{p1}{\footnotesize{$+1$}}
\psfrag{C1m}{\footnotesize{$C_{1}^{-}$}}
\psfrag{C1p}{\footnotesize{$C_{1}^{+}$}}
\psfrag{C1k}{\footnotesize{$C_{K/2}^{-}$}}
\psfrag{C1j}{\footnotesize{$C_{K/2}^{+}$}}
\psfrag{a}{\footnotesize{$x^{(K)}_{-}$}}
\psfrag{b}{\footnotesize{$x^{(K-1)}_{-}$}}
\psfrag{c}{\footnotesize{$x^{(2)}_{-}$}}
\psfrag{d}{\footnotesize{$x^{(1)}_{-}$}}
\psfrag{e}{\footnotesize{$x^{(1)}_{+}$}} 
\psfrag{f}{\footnotesize{$x^{(2)}_{+}$}}
\psfrag{g}{\footnotesize{$x^{(K-1)}_{+}$}}
\psfrag{h}{\footnotesize{$x^{(K)}_{+}$}}
\includegraphics[width=100mm]{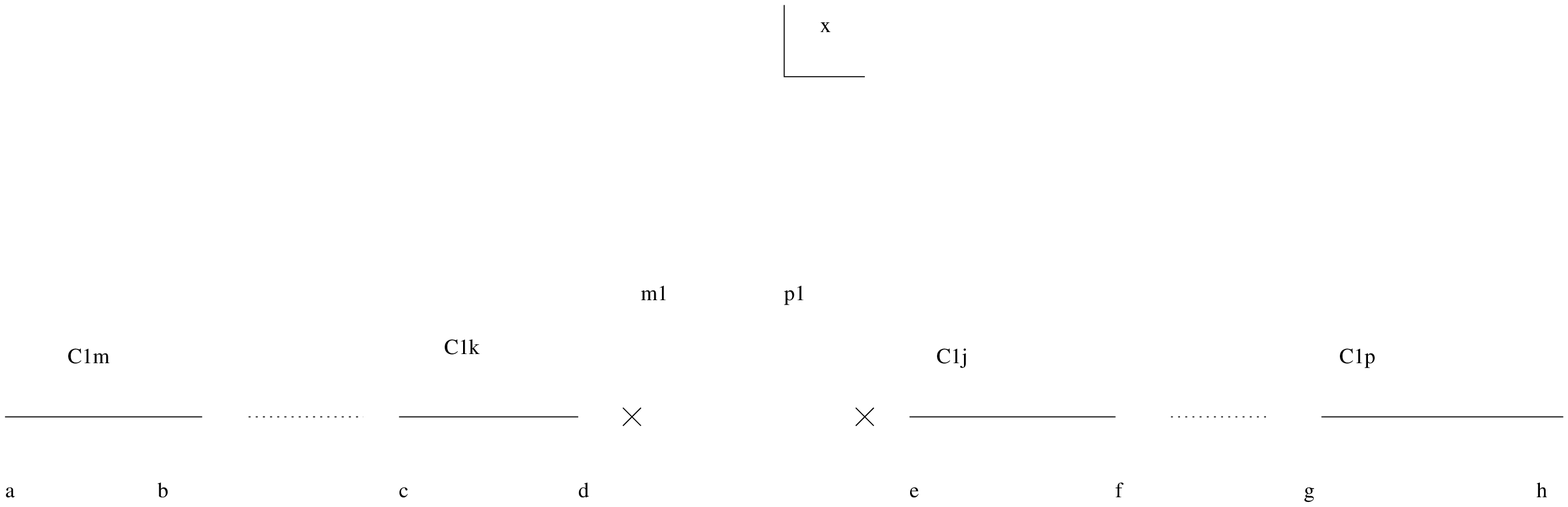}
\caption{The cut $x$-plane corresponding to the curve $\Sigma$}
\label{Sfig1}
\end{figure}

\begin{figure}
\centering
\psfrag{AI}{\footnotesize{$\mathcal{A}^{\pm}_{I}$}}
\psfrag{BI}{\footnotesize{$\mathcal{B}^{\pm}_{I}$}}
\psfrag{CI}{\footnotesize{$C^{\pm}_{I}$}}
\psfrag{Ip}{\footnotesize{$\infty^{+}$}}
\psfrag{Im}{\footnotesize{$\infty^{-}$}}
\includegraphics[width=100mm]{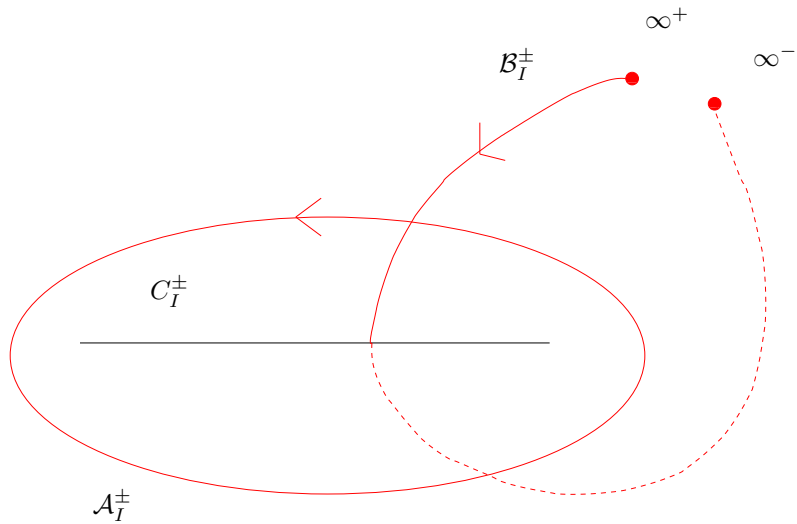}
\caption{The cycles on $\Sigma$. The
  index $I$ runs from $1$ to $K/2$.}
\label{Sfig2}
\end{figure}

The construction of solutions to the string equations of
motion for this system is given in \cite{KZ} and reviewed in
\cite{DS}. Solutions are labelled by the number $K\geq 0$ of gaps in
the spectrum of the auxiliary linear problem. The $K$-gap solutions
are characterised by a hyperelliptic curve $\Sigma$ of genus $K-1$ and
a meromorphic differential $dp$ on $\Sigma$ with prescribed
singularities and asymptotics which will be reviewed below. For
simplicity we will focus on the case of even $K$. 
\paragraph{}
 In the following we will also focus on curves where all the branch points
lie on the real axis and outside the interval $[-1,+1]$. This
corresponds to string solutions where only classical oscillator modes
which carry positive spin are activated. In other words, only the
transverse modes of the string with angular momentum $S=+1$ are
excited. 
In the dual gauge theory
these solutions are believed to correspond to operators of the form 
(\ref{sl2}) where only the covariant derivative 
$\mathcal{D}_{+}$, which carries positive spin, appears \cite{KZ}. 
\paragraph{}
We label the branch points of the curve according to,  
\bea 
\Sigma\,\,:\qquad{} y^{2} & = & \prod_{i=1}^{K} 
\left(x-x^{(i)}_{+}\right)\left(x-x^{(i)}_{-}\right)
\label{curve} \eea
with the ordering, 
\bea 
x_{-}^{(K)}\,\,\leq \,\,x_{-}^{(K-1)} & \ldots &\leq
x_{-}^{(1)}\,\,\leq -1  \nn \\ 
x_{+}^{(K)}\,\,\geq \,\,x_{+}^{(K-1)} & \ldots &\geq
x_{+}^{(1)}\,\,\geq +1 \nn \eea 
The branch points are joined in pairs by cuts $C^{\pm}_{I}$, 
$I=1,2,\ldots,K/2$ as shown in Figure \ref{Sfig1}.  
We also define a standard basis of one-cycles,
$\mathcal{A}^{\pm}_{I}$, $\mathcal{B}^{\pm}_{I}$. Here
$\mathcal{A}^{\pm}_{I}$ encircles the cut $C^{\pm}_{I}$ on the upper
sheet in an anti-clockwise direction and $\mathcal{B}^{\pm}_{I}$ runs
from the point at infinity on the upper sheet to the point at infinity
on the lower sheet passing through the cut  $C^{\pm}_{I}$, as shown in
Figure \ref{Sfig2}.
\paragraph{}
The meromorphic differential $dp$ on $\Sigma$ 
has second-order poles at the points above $x=+1$ and
$x=-1$ on $\Sigma$ whose coefficients encode the $S^{1}$ angular
momentum of the solution. On the top sheet we have, 
%\bea 
%dp & \longrightarrow & 
%\frac{\pi \kappa_{\pm}\,dx}{(x\pm
%    1)^{2}}\,\, + \,\, O\left((x\pm 1)^{0}\right)  
%\label{poles} 
%\eea 
%as $x\rightarrow \mp 1$. Here, 
%\bea 
%\kappa_{\pm} & = & \frac{J}{\sqrt{\lambda}}\,\,\pm \,\,m \nn \eea 
%where $m\in \mathbb{Z}$ is related to the total worldsheet momentum as
%$P=2\pi m$. 
\bea 
dp & \longrightarrow & -
 \frac{\pi J}{\sqrt{\lambda}} \frac{dx}{(x\pm
    1)^{2}}\,\, + \,\, O\left((x\pm 1)^{0}\right)  
\label{poles} 
\eea
as $x\rightarrow \mp 1$.
\paragraph{}
There are also two second-order poles at the
points $x=\mp 1$ on the lower sheet related by the involution 
$dp\rightarrow -dp$. The value of the Noether charges $\Delta$ and $S$
is encoded in the asymptotic behaviour of $dp$ near the points $x=0$
and $x=\infty$ on the top sheet, 
\bea
dp & \longrightarrow & -\frac{2\pi}{\sqrt{\lambda}}\,\,
\left(\Delta\,+\,S \right)\,\,\frac{dx}{x^{2}} \qquad{} {\rm as}\,\,\,\, 
x\rightarrow \infty \label{asymp1} \\ 
dp & \longrightarrow & -\frac{2\pi}{\sqrt{\lambda}}\,\,
\left(\Delta\,-\,S\right)\,\,dx \qquad{} {\rm
  as}\,\,\,\, 
x\rightarrow 0 \label{asymp2} \eea
For a valid semiclassical description, the conserved charges $J$, $S$ and
$\Delta$ should all be $O(\sqrt{\lambda})$ with $\sqrt{\lambda}>>1$.  
\paragraph{}
In addition to the above relations, the closed string boundary
condition imposes $2K$ normalisation conditions on $dp$, 
\bea 
\oint_{\mathcal{A}^{\pm}_{I}} \,dp\,\,=\,\,0   & \qquad{} & 
\oint_{\mathcal{B}^{\pm}_{I}} \,dp \,\,=\,\,2\pi n^{\pm}_{I} 
\label{norm} \eea 
with $I=1,2,\ldots, K/2$. The integers $n^{\pm}_{I}$ correspond to the
mode numbers of the string. In the following we 
will assign the mode numbers so as to
pick out the $K$ lowest modes of the string which carry positive
angular momentum, 
including both left and
right movers. This is accomplished by setting $n^{\pm}_{I}=\pm I$ for
$I=1,\ldots, K/2$. 
\paragraph{}
To find the spectrum of classical string solutions
we must first construct the 
meromorphic differential $dp$ with the specified pole
behaviour (\ref{poles}). The most general possible such differential
has the form, 
\bea 
dp  & = & dp_{1}\,\,+\,\,dp_{2}\,\,=\,\,-\frac{dx}{y}\left[f(x)\,\,+\,\,
g(x)\right]  \label{ansatz} \\ 
f(x) & = & \sum_{\ell=0}^{K-2}\,\,C_{\ell}\,x^{\ell} \nn \\
g(x) & = & \frac{\pi J}{\sqrt{\lambda}}\left[
\frac{y_{+}}{(x-1)^{2}}\,\,+\,\,\frac{y_{-}}{(x+1)^{2}} 
+\frac{y'_{+}}{(x-1)}\,\,+\,\,\frac{y'_{-}}{(x+1)}\right] \nn \eea 
with $y_{\pm}=y(\pm 1)$ and 
\bea 
y'_{\pm} & = & \left. \frac{dy}{dx}\right|_{x=\pm 1} \nn \eea  
Here the second term $dp_{2}$ is a particular differential with the
required poles and the first term $dp_{1}$ is a general holomorphic
differential on $\Sigma$. The resulting curve $\Sigma$ and
differential $dp$ depend on
$3K-1$ undetermined parameters $\{x^{(i)}_{\pm}, C_{\ell}\}$
with $i=1,\ldots, K$, $\ell=0,1,\ldots,K-2$. We then obtain $2K$
constraints on these parameters from the normalisation equations
(\ref{norm}), leaving us with a $K-1$ dimensional moduli space of
solutions \cite{KZ}. %As mentioned above, 
A significant difficulty with this
approach is that the normalisation conditions are transcendental and
cannot be solved in closed form. 
\paragraph{} 
A convenient parametrisation for the moduli space is given in terms of
the $K$ {\em filling fractions}, 
\bea 
\mathcal{S}^{\pm}_{I} & = & - \frac{1}{2\pi i}\,\cdot\,
\frac{\sqrt{\lambda}}{4\pi}\,\,\oint_{\mathcal{A}^{\pm}_{I}}\,
\left(x\,+\,\frac{1}{x}\right)\,dp
\label{eq:filling_fractions}
\eea 
with $I=1,\ldots,K/2$, subject to the level matching constraint, 
%\bea 
%\sum_{I=1}^{K/2}\,\,n^{+}_{I}\mathcal{S}^{+}_{I}\,\,+\,\,
%n^{-}_{I}\mathcal{S}^{-}_{I} & = & P  \label{level} \eea
%where $P$ is the total worldsheet momentum. 
\bea 
\sum_{I=1}^{K/2}\,\, \left( n^{+}_{I}\mathcal{S}^{+}_{I}\,\,+\,\,
n^{-}_{I}\mathcal{S}^{-}_{I} \right) & = & 0  \label{level} \eea
Here the total $AdS$ angular momentum is given as 
\bea 
S & = & \sum_{I=1}^{K/2}\, \left( \mathcal{S}^{+}_{I}\,\,+\,\,
\mathcal{S}^{-}_{I} \right) \nn \eea
and is regarded as one of the moduli of the solution. The significance
of the filling fractions is that they constitute a set of 
normalised action variables for the string\footnote{
The symplectic structure of the string was analysed in detail for the case of
  strings on $S^{3}\times \mathbb{R}$ in \cite{DV1,DV2}. The resulting
  string $\sigma$-model was an $SU(2)$ principal chiral model (PCM). In the
  context of the finite gap construction one works with a complexified
  Lax connection and results for the $SU(2)$ 
and $SL(2,\mathbb{R})$ PCMs differ only at the level of reality
  conditions which do not affect the conclusion that the
filling fractions are the canonical action variables of the string.}. 
They are canonically
conjugate to angles $\varphi_{I}\in [0,2\pi]$ living on the Jacobian
torus $\mathcal{J}(\Sigma)$. Evolution of the string solution in both
worldsheet coordinates, $\sigma$ and $\tau$, corresponds to linear
motion of these angles \cite{DV1}.  
\paragraph{}
The constraints described above uniquely determine $(\Sigma,dp)$ for
given values of $\mathcal{S}^{\pm}_{I}$, and 
one may then extract the 
string energy from the asymptotics (\ref{asymp1},\ref{asymp2}) which imply,  
%\bea 
%\Delta\,+\,S & = & -\frac{\sqrt{\lambda}}{2\pi}\, C_{K-2} \nn \\ 
%\Delta\,-\,S & = & -\frac{\sqrt{\lambda}}{2\pi}\, \frac{C_{0}}{y(0)}
%\,\,+ \,\, \frac{J}{2y(0)}\left(y_{+}+y_{-}-y'_{+}-y'_{-}\right) \nn 
%\eea
\bea 
\Delta\,+\,S & = & \frac{\sqrt{\lambda}}{2\pi}\, C_{K-2} \nn \\ 
\Delta\,-\,S & = & \frac{\sqrt{\lambda}}{2\pi}\, \frac{C_{0}}{y(0)}
\,\,+ \,\, \frac{J}{2y(0)}\left(y_{+}+y_{-}-y'_{+}+y'_{-}\right) \nn 
\eea
In this way, one obtains a set of transcendental equations which determine
the string energy as a function of the filling fractions,  
\bea
\Delta & = & \Delta\left[\mathcal{S}^{+}_{1},\mathcal{S}^{-}_{1}, \ldots, 
\mathcal{S}^{+}_{K/2},\mathcal{S}^{-}_{K/2}\right] \nn \eea 
The leading order semiclassical spectrum of the string is
obtained by imposing the Bohr-Sommerfeld conditions which impose the
integrality of the filling fractions \cite{DV1, DV2}: $\mathcal{S}^{\pm}_{I}\in
\mathbb{Z}$, $I=1,2,\ldots, K/2$. 
\paragraph{}
Finally we define the quasi-momentum as the abelian integral of the
differential $dp$, 
\bea 
p(x) & = & \int_{\infty^{+}}^{x}\,\,dp 
\eea
where $\infty^+$ indicates complex infinity on the upper sheet.
%The worldsheet momentum can be extracted from the quasi-momentum via
%the formula \cite{KZ}, 
%\bea 
%P\,\,=\,\,2\pi m & = & \frac{1}{2\pi i} \,\,\sum_{I=1}^{K/2}\,\,\left[ 
%\oint_{\mathcal{A}^{+}_{I}}\,\,+\,\,
%\oint_{\mathcal{A}^{-}_{I}}\right] \,\,\frac{p(x)\,dx}{x} \nn \eea
%The total worldsheet momentum is related to the filling fractions via
%the level matching constraint (\ref{level}). 

\section{The large spin limit}
\label{sec:large_S}
\paragraph{}
Our goal now is to investigate the large spin limit of the finite gap
construction and in particular focus on excitations around the long
spinning string solution. Here we will repeat and extend the analysis of
\cite{DS}. To this end we label the branch points in the following
way, 
\bea 
x^{(i)}_{\pm} & = & b \qquad{} \qquad{}\,\,\, i=1 \nn \\ & = & 
b^{(i-1)}_{\pm} \qquad{} \,\,\,\,\,\,\,\, i=2,\ldots,M+1 \nn \\ & = & 
a^{(i-M-1)}_{\pm} \qquad{} \,\,\, i=M+2,\ldots,K 
\nn \eea
where $M$ must be \emph{even}.
To obtain large spin, it is necessary to scale the coordinates of at
least two of the branch points linearly with $S$ as $S\rightarrow
\infty$. Here we will take a generic limit of this kind where the
``$a$'' branch points introduced above scale with $S$ while the ``$b$''
branch points are held fixed. We introduce a scaling parameter 
$\rho$ and set  
\bea a^{(j)}_{\pm} & = & \rho \tilde{a}_{\pm}^{(j)} \nn \eea
for $j=1,2,\ldots, K-M-1$ and take the limit $\rho \rightarrow \infty$
 with $\tilde{a}^{(j)}_{\pm}$,  $b^{(j)}_{\pm}$ and $b$ held
 fixed. Thus we are dividing the branch points into ``large'' and ``small''.
This is a generalisation of the limit considered in \cite{DS}
(which in the present notation corresponds to $M = 0$) and we
will draw heavily on the results of this reference. 
\paragraph{}
The key point is that the limit must be taken in such a way that the
conditions on the periods of the differential $dp$ are preserved. As
explained in \cite{DS} the scaling leads to a very 
specific degeneration of the
spectral curve $\Sigma$ into two components;
\bea 
\Sigma &\longrightarrow& \tilde{\Sigma}_{1} \cup \tilde{\Sigma}_{2}
\nn \eea
\paragraph{}
The two components $\tilde{\Sigma}_{1}$ and $\tilde{\Sigma}_{2}$
correspond to the ``large'' and ``small'' branch points respectively. 
The curves are given explicitly as 
\bea 
\tilde{\Sigma}_{1}\,\,:\qquad{} \tilde{y}^{2}_{1} & = & \prod_{i=1}^{K-M-1} 
\left(\tilde{x}-\tilde{a}^{(i)}_{+}\right)\left(\tilde{x}-
\tilde{a}^{(i)}_{-}\right) \:,
\nn \eea
where $\tilde{x} = x / \rho$ is the rescaled spectral parameter, and 
\bea 
\tilde{\Sigma}_{2}\,\,:\qquad{} \tilde{y}^{2}_{2} & = &
(x^{2}\,-\,b^{2}) \prod_{i=1}^{M}\left(x-b^{(i)}_{+}\right) 
\left(x-b^{(i)}_{-}\right)
\nn \eea
These two surfaces have genus $K-M-2$ and $M$ respectively. At this
point we have not yet imposed the period conditions (\ref{norm}).
In the
following we will see that these two surfaces describe the dynamics of
``large'' and ``small'' spikes respectively. The differential $dp$
also decomposes into differentials $d\tilde{p}_{1}$ and
$d\tilde{p}_{2}$ on each of these curves, 
\bea  
dp & \longrightarrow  d\tilde{p}_{1}\,\,\,\,+\,\,\,\,
d\tilde{p}_{2} \nn \eea

\subsection*{Review of the simple case $M=0$}

\begin{figure}
\centering
\psfrag{S1}{\footnotesize{$\Sigma$}}
\psfrag{S2}{\footnotesize{$\tilde{\Sigma}_{2}$}}
%\psfrag{B}{\footnotesize{$\hat{\mathcal{B}}$}} 
%\psfrag{A2}{\footnotesize{$\hat{\mathcal{A}}_{2}$}}
\psfrag{Ip}{\footnotesize{$\infty^{+}$}}
\psfrag{Im}{\footnotesize{$\infty^{-}$}}
\includegraphics[width=100mm]{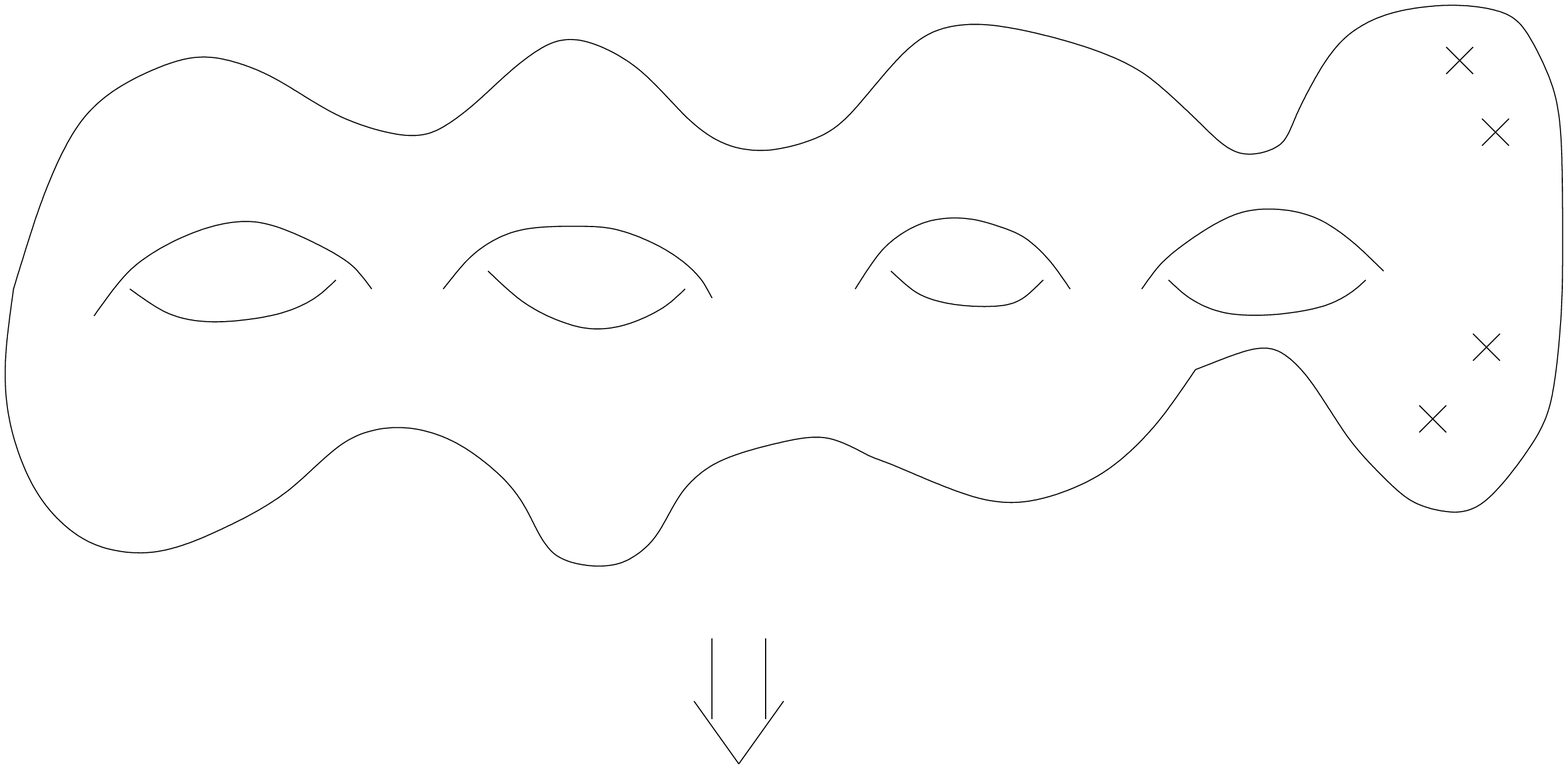}
%\caption{The ``extra'' cycle $\mathcal{A}^{+}_{K/2}$ becomes the sum
%  of the chains $\hat{\mathcal{A}}_{1}$ and $\hat{\mathcal{A}}_{2}$.}
%\label{Sfig3b}
%\end{figure}
%\begin{figure}
\centering
%\psfrag{B}{\footnotesize{${\mathcal{C}}$}} 
\psfrag{S1}{\footnotesize{$\tilde{\Sigma}_{1}$}}
\psfrag{S2}{\footnotesize{$\tilde{\Sigma}_{2}$}}
%\psfrag{CI}{\footnotesize{$C^{\pm}_{I}$}}
%\psfrag{Ip}{\footnotesize{$\infty^{+}$}}
%\psfrag{Im}{\footnotesize{$\infty^{-}$}}
\includegraphics[width=100mm]{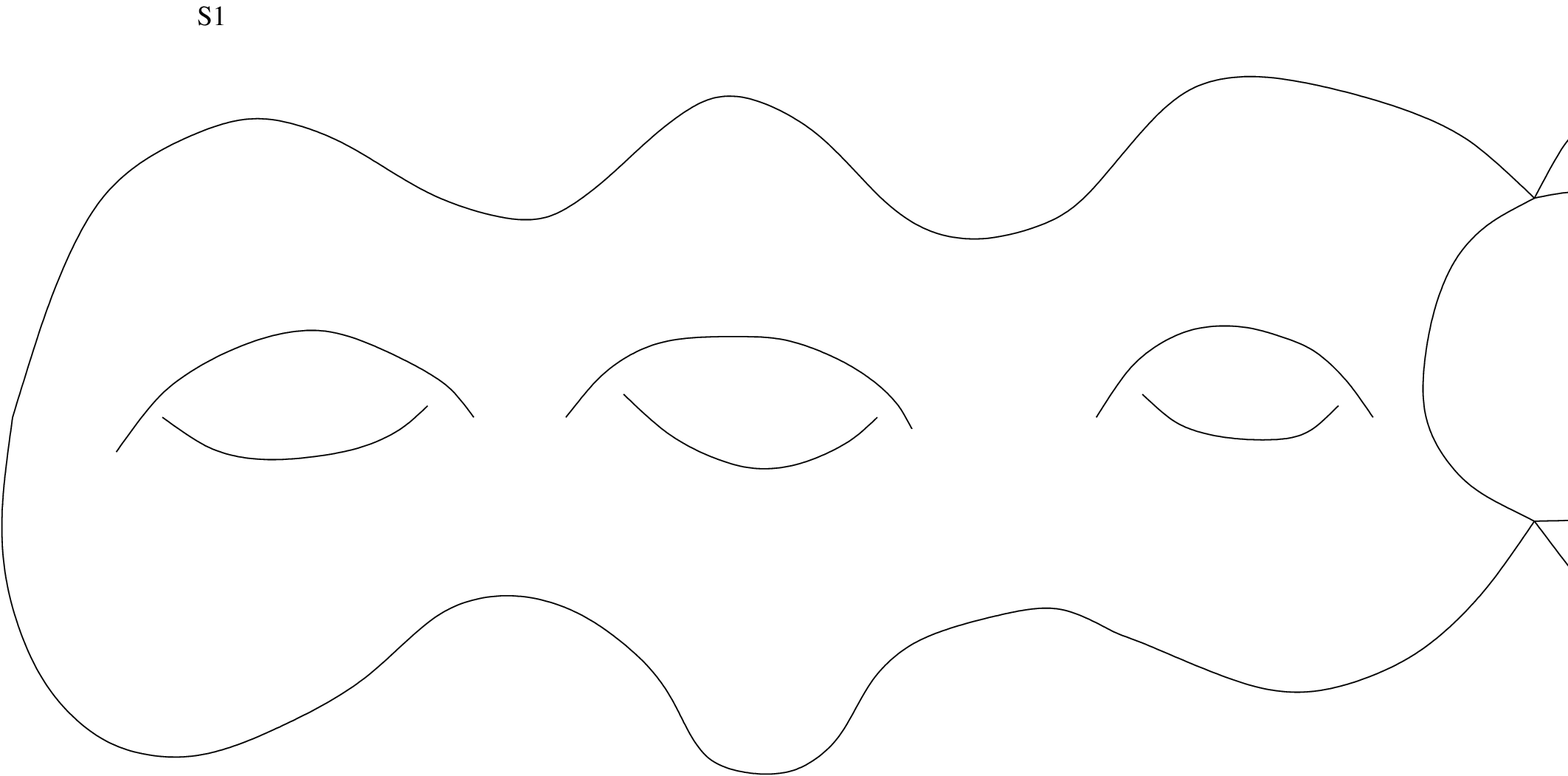}
\caption{The degeneration $\Sigma \rightarrow \tilde{\Sigma}_{1}\cup
\tilde{\Sigma}_{2}$. The four singular points $\pm 1^{\pm}$ are marked
with crosses on the curve.}
\label{Sfig3}
\end{figure}

We begin by briefly reviewing the case considered in \cite{DS} which 
corresponds to $M=0$. The corresponding degeneration of the spectral
curve is illustrated in Figure \ref{Sfig3}. 
The key point is that the differential $d\tilde{p}_{1}$ on
$\tilde{\Sigma}_{1}$ has only simple poles and it is then possible to
solve the conditions on the periods of this differential 
to find an explict parametrization
of the curve.    
The relevant formulae are, 
\begin{equation}
 \tilde{\Sigma}_1 \:\: : \qquad \tilde{y}_1^2 = \prod_{i = 1}^{K-1} \left( \tilde{x} - \tilde{a}^{(i)}_+ \right) \left( \tilde{x} - \tilde{a}^{(i)}_- \right)
\label{eq:Sigma_1_tilde_M=0}
\end{equation}
and
\begin{equation}
 d \tilde{p}_1 = -i \frac{d \tilde{x}}{\tilde{x}^2} \frac{\mathbb{P}'_K \left( \frac{1}{\tilde{x}} \right)}{\sqrt{\mathbb{P}_K^2
  \left( \frac{1}{\tilde{x}} \right) -4}}
\label{eq:dp1tilde_M=0}
\end{equation}
with
\begin{equation}
 \mathbb{P}_K \left( \frac{1}{\tilde{x}} \right) = 2 + \frac{\tilde{q}_2}{\tilde{x}^2} + \frac{\tilde{q}_3}{\tilde{x}^3} +
  \ldots + \frac{\tilde{q}_K}{\tilde{x}^K} \:,
\label{eq:P_K(1/x)_M=0}
\end{equation}
The curve coincides with the curve of a classical $SL(2,\mathbb{R})$
spin chain of length $K$ and $d\tilde{p}_{1}$ is precisely the
quasi-momentum of this system. The $K-1$ remaining moduli
$\tilde{q}_{i}$ are the conserved charges of the spin chain.
\paragraph{}
It is also straightforward to find the limiting form of the second
component and the corresponding differential, 
\bea 
\tilde{\Sigma}_{2}\,\,:\qquad{} \tilde{y}^{2}_{2} & = &
(x^{2}\,-\,b^{2}) \nn \eea 
and 
\bea 
d\tilde{p}_{2} & = & -\,\frac{i\pi J}{\sqrt{\lambda}}\left[ 
\sqrt{b^{2}-1}\left(\frac{1}{(x-1)^{2}}\,+\,\frac{1}{(x+1)^{2}}\right)
\,\,-\,\, 
\frac{1}{\sqrt{b^{2}-1}}\left(\frac{1}{(x-1)}\,-\,\frac{1}{(x+1)}\right)
\,\right]\,\,\frac{dx}{\tilde{y}_{2}}\nn  \\
& & \qquad - \frac{K}{i} \frac{dx}{\tilde{y}_2}
\label{dp2} \eea
\begin{figure}
\centering
\psfrag{S1}{\footnotesize{$\Sigma$}}
\psfrag{S2}{\footnotesize{$\tilde{\Sigma}_{2}$}}
%\psfrag{Ah}{\footnotesize{$\mathcal{A}^{+}_{K/2}$}} 
%\psfrag{A2}{\footnotesize{$\hat{\mathcal{A}}_{2}$}}
%\psfrag{Ip}{\footnotesize{$\infty^{+}$}}
%\psfrag{Im}{\footnotesize{$\infty^{-}$}}
\includegraphics[width=100mm]{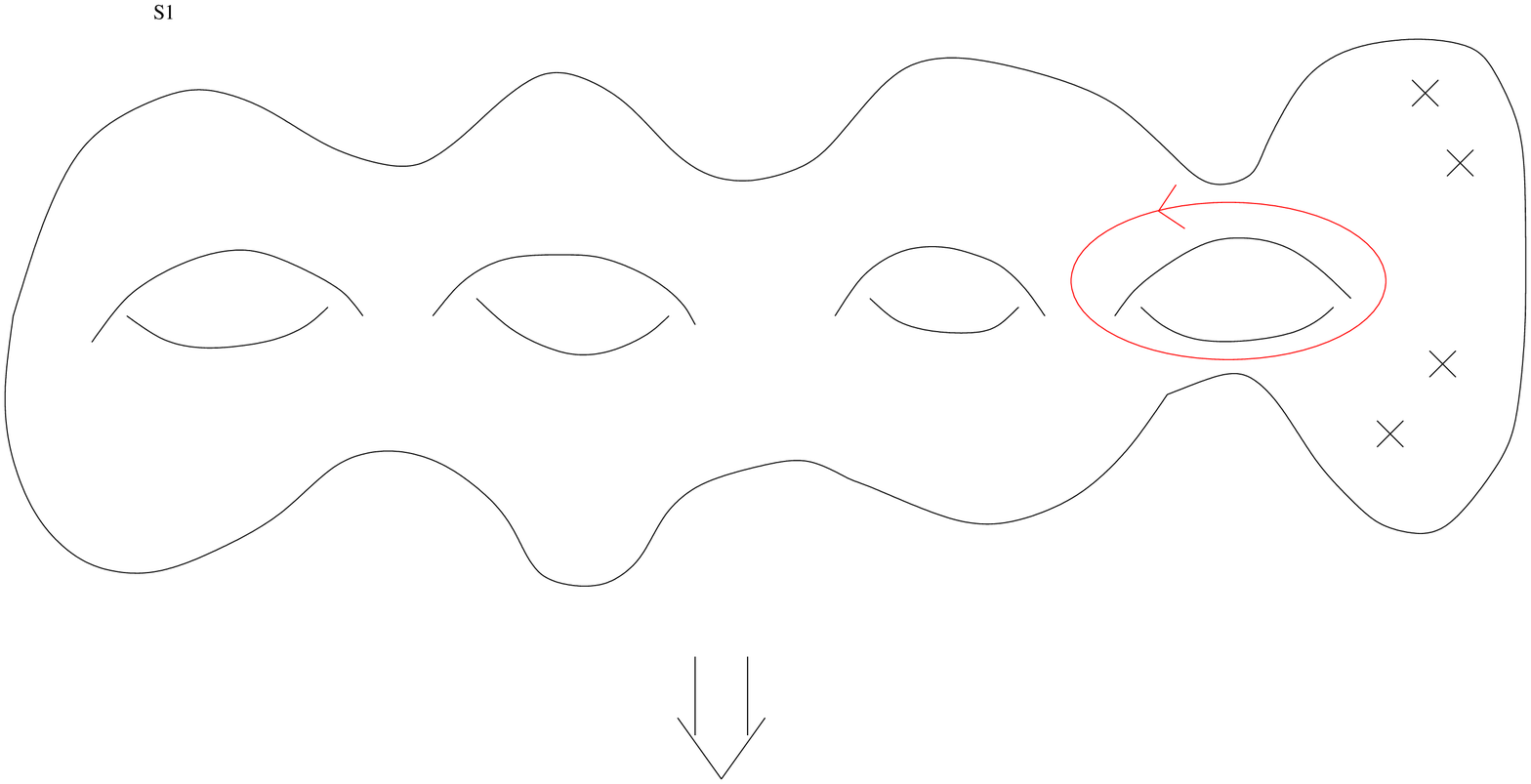}
%\caption{The ``extra'' cycle $\mathcal{A}^{+}_{K/2}$ becomes the sum
%  of the chains $\hat{\mathcal{A}}_{1}$ and $\hat{\mat.}
\label{Sfig6b}
%\end{figure}
%\begin{figure}
\centering
\psfrag{S1}{\footnotesize{$\tilde{\Sigma}_{1}$}}
\psfrag{S2}{\footnotesize{$\tilde{\Sigma}_{2}$}}
%\psfrag{A1}{\footnotesize{$\hat{\mathcal{A}}_{1}$}} 
%\psfrag{A2}{\footnotesize{$\hat{\mathcal{A}}_{2}$}}
%\psfrag{Ip}{\footnotesize{$\infty^{+}$}}
%\psfrag{Im}{\footnotesize{$\infty^{-}$}}
\includegraphics[width=100mm]{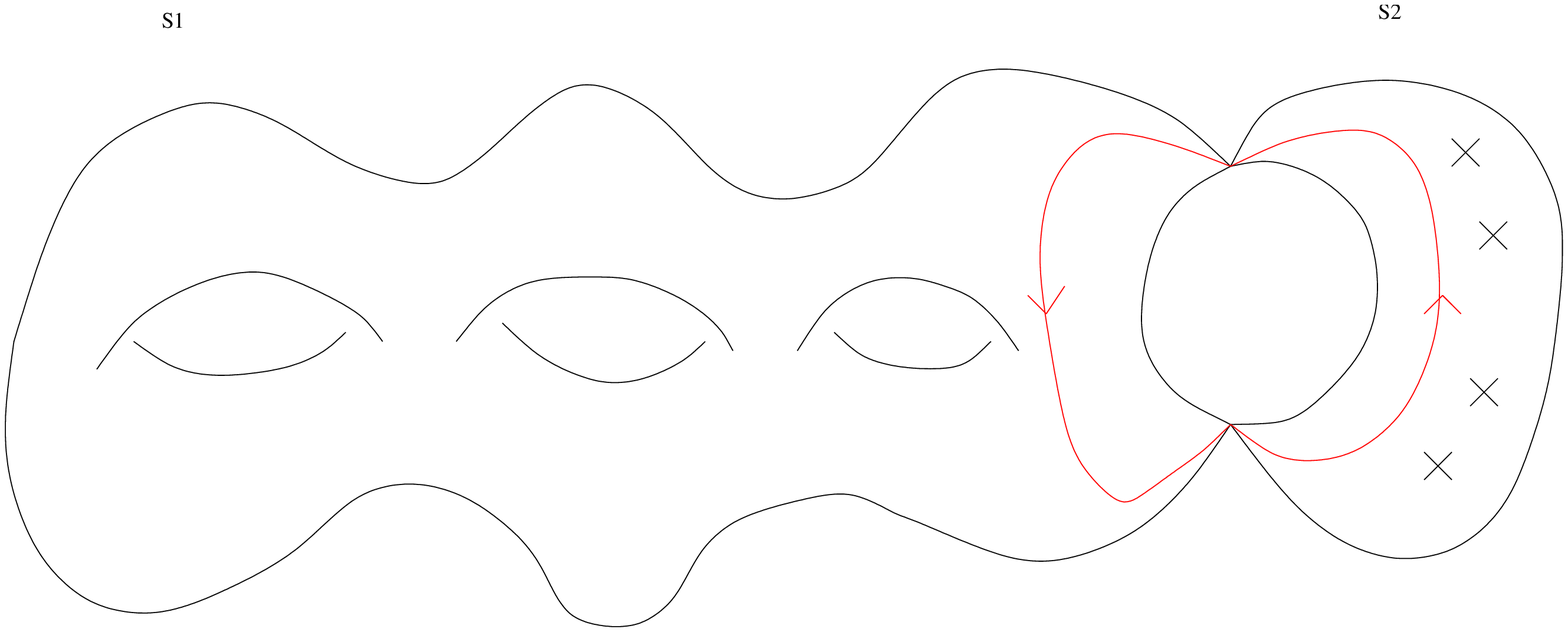}
\caption{The ``extra'' cycle $\mathcal{A}^{+}_{K/2}$ gives rise to the
  matching condition.}
\label{Sfig6}
\end{figure}
Of the period conditions on the original surface $\Sigma$, all but one
become period conditions on the differential $d\tilde{p}_{1}$ which
are solved explicitly in (\ref{eq:dp1tilde_M=0}). The ``extra'' period
becomes a sum of chains on both components. The corresponding
normalisation equation gives rise to a matching condition which
relates the moduli of the two component surfaces and takes the form,      
\bea \sqrt{b^{2}-1} & = & \frac{2\pi J}{\sqrt{\lambda}}\,\times \, 
\frac{1}{  K\log \left(\rho \tilde{q}_{K}^{1/K}\right)} \label{match2}
\eea 
\paragraph{}
In the $\rho\rightarrow \infty$ limit, 
the conserved charges have the behaviour, 
\bea 
\Delta\,+\,S & \simeq & \frac{\sqrt{\lambda}}{2\pi} \sqrt{- \tilde{q}_2} \,\,\rho 
\qquad{} \rightarrow 
\,\,\,\,\infty \nn \\ 
\Delta\,-\,S & \simeq & - \frac{\sqrt{\lambda}}{2\pi} \,\, \frac{K}{b} 
\,\,\,+\,\,\,\frac{J}{b}\left(\sqrt{b^{2}-1}\,+\,
\frac{1}{\sqrt{b^{2}-1}}\right) \label{behave} \eea
up to corrections which vanish as $\rho\rightarrow \infty$. 
The behaviour of the inner branch points $x=\pm b$ in this limit 
is determined by
the matching condition (\ref{match2}).  
Thus, as the scaling parameter $\rho$ goes to infinity, $b\rightarrow 1$ and 
the inner branch points approach the punctures at the points
$x=\pm 1$. 
Comparing (\ref{match2}) with (\ref{behave}) we find the limiting
behaviour, 
\bea  \sqrt{b^{2}-1} & \simeq & \frac{J}{\Delta-S} \nn \eea 
Finally taking account of this limit in (\ref{dp2}) we find the
limiting form of the quasi-momentum, 
\bea
\left. d\tilde{p}_{2}\right |_{M=0} & \simeq & \frac{2\pi i}{\sqrt{\lambda}}\, (\Delta -S)
\,\, \frac{dx}{(x^{2}-1)^{\frac{3}{2}}}
\label{dp2b}
\eea
This is the leading contribution to the quasi-momentum in the limit
$S\rightarrow \infty$ up to contributions which are supressed by
inverse powers of $\log S$, while the anomalous dimension is given by:
\begin{equation}
 \Delta - S \simeq \frac{\sqrt{\lambda}}{2 \pi} \left[ K \log S + \log \left( \frac{\tilde{q}_K}{(- \tilde{q}_2)^{K/2}} \right) + \mathrm{const.} \right] + \ldots
\label{eq:Delta-S_as_f-n_of_the_moduli}
\end{equation}
where the constant is moduli-independent. This moduli-dependent term
in this expression coincides with the nearest-neighbour Hamiltonian of
the $SL(2,\mathbb{R})$ spin chain. 
\paragraph{}
The main result of the analysis of \cite{DS} is that
the semiclassical spectrum of string theory on $AdS_{3}\times S^{1}$
in the particular limit described above is captured by the classical
spin chain. In \cite{DL1}, approximate solutions corresponding to
this spectrum were constructed. The solutions include $K$ large spikes
with angular positions which can be related to the moduli
$\tilde{q}_{i}$ appearing in the spin chain curve. Our next objective
is to generalise this analysis to the more general limit discussed
above with $M>0$. Thus we will include ``small'' branch points and we
will see that they correspond to the contributions of small spikes.

\subsection*{The general case $M$ even}

The main difference from the previous case is the fact that now $2M + 2$ ``small'' branch points (located at $x = b^{(j)}_\pm$, $j = 1, \ldots, M$, and at the usual $x = \pm b$) migrate to $\tilde{\Sigma}_2$ as $\rho \to \infty$. Thus, as explained above, after the factorisation this surface has, in principle, genus $M$, although we will later see that this will reduce to $1$ due to cut collisions. $\tilde{\Sigma}_1$, on the other hand, has now genus $K - M - 2$ and contains all the ``large'' branch points. The extra cuts transferred to $\tilde{\Sigma}_2$ make the analysis more difficult on that surface.

\begin{figure}
\centering
\psfrag{a}{\footnotesize{(a)}}
\psfrag{b}{\footnotesize{(b)}}
\psfrag{c}{\footnotesize{(c)}}
\psfrag{a1}{\footnotesize{$\mathcal{A}_{\frac{K-M}{2}-1}^+$}}
\psfrag{a2}{\footnotesize{$\mathcal{A}_1^+$}}
\psfrag{a3}{\footnotesize{$\mathcal{A}_{\frac{K-M}{2}-1}^-$}}
\psfrag{a4}{\footnotesize{$\mathcal{A}_1^-$}}
\psfrag{a5}{\footnotesize{$\tilde{\mathcal{A}}_0$}}
\psfrag{a6}{\footnotesize{$\hat{\mathcal{A}}_{\frac{K}{2}-1}^-$}}
\psfrag{a7}{\footnotesize{$\hat{\mathcal{A}}_{\frac{K}{2}}^-$}}
\psfrag{a8}{\footnotesize{$\hat{\mathcal{A}}_{\frac{K-M}{2}}^+$}}
\psfrag{a9}{\footnotesize{$\hat{\mathcal{A}}_{\frac{K}{2}-1}^+$}}
\psfrag{a10}{\footnotesize{$\hat{\mathcal{A}}_{\frac{K}{2}}^+$}}
\includegraphics[width=\columnwidth]{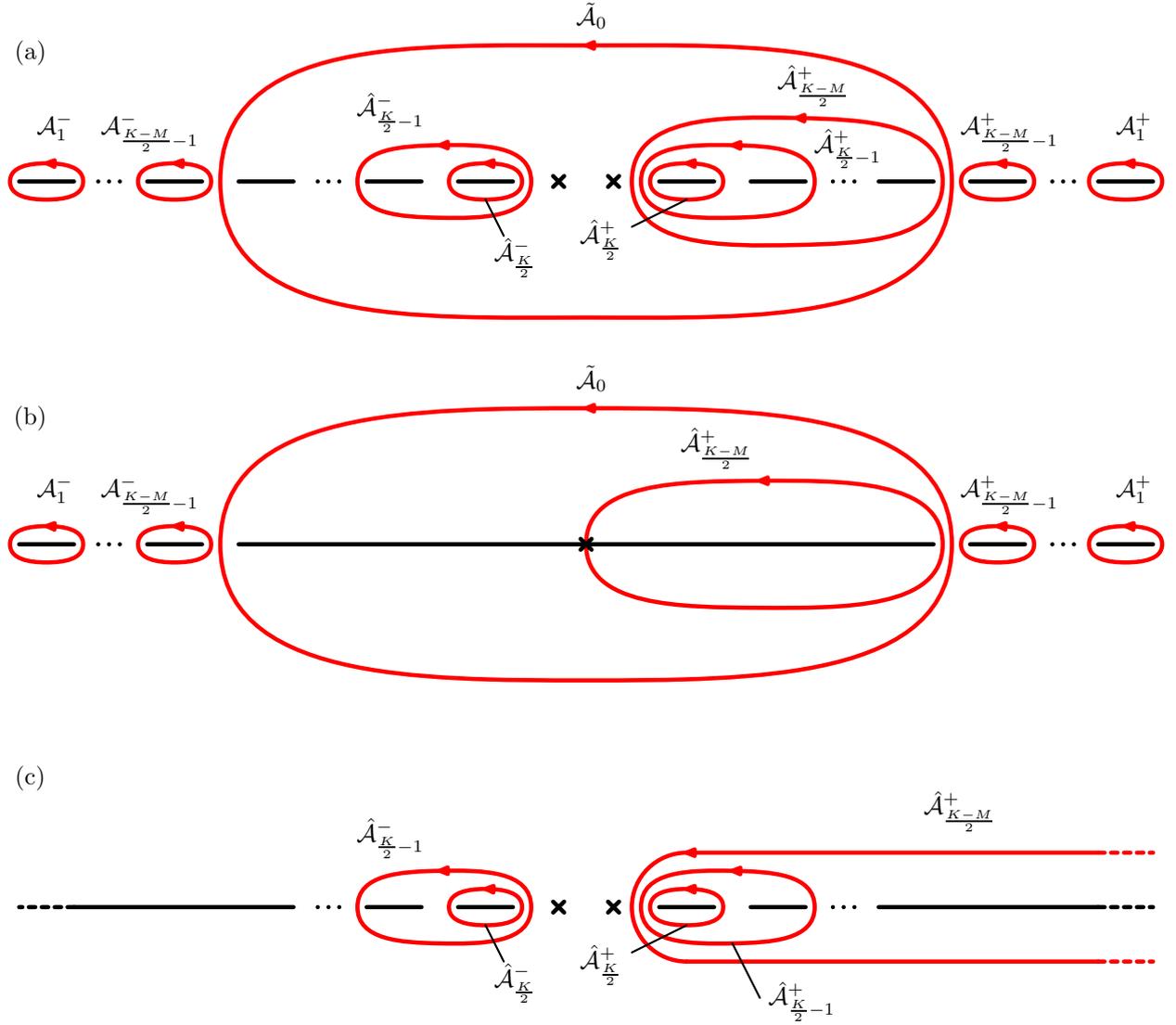}%
\caption{(a) The rearranged A-cycle configuration on $\Sigma$. (b) The A-cycles on $\tilde{\Sigma}_1$; the marked point indicates the simple pole at $x=0$. (c) The A-cycles on $\tilde{\Sigma}_2$; the marked points indicate the double poles at $x = \pm 1$. All dashed lines extend to infinity along the real axis.}
\label{fig:Acycles}%
\end{figure}

\begin{figure}
\centering
\psfrag{a}{\footnotesize{(a)}}
\psfrag{b}{\footnotesize{(b)}}
\psfrag{c}{\footnotesize{(c)}}
\psfrag{b1}{\footnotesize{$\mathcal{B}_1^-$}}
\psfrag{b2}{\footnotesize{$\mathcal{B}_{\frac{K-M}{2}-1}^-$}}
\psfrag{b3}{\footnotesize{$\mathcal{B}_1^+$}}
\psfrag{b4}{\footnotesize{$\mathcal{B}_{\frac{K-M}{2}-1}^+$}}
\psfrag{b5}{\footnotesize{$\mathcal{B}_{\frac{K-M}{2}}^+$}}
\psfrag{b6}{\footnotesize{$\hat{\mathcal{B}}$}}
\psfrag{b7}{\footnotesize{$\hat{\mathcal{B}}_{\frac{K}{2}}^-$}}
\psfrag{b8}{\footnotesize{$\hat{\mathcal{B}}_{\frac{K}{2}-1}^-$}}
\psfrag{b9}{\footnotesize{$\hat{\mathcal{B}}_{\frac{K-M}{2}+1}^-$}}
\psfrag{b10}{\footnotesize{$\hat{\mathcal{B}}_{\frac{K}{2}}^+$}}
\psfrag{b11}{\footnotesize{$\hat{\mathcal{B}}_{\frac{K}{2}-1}^+$}}
\psfrag{b12}{\footnotesize{$\hat{\mathcal{B}}_{\frac{K-M}{2}+1}^+$}}
\includegraphics[width=\columnwidth]{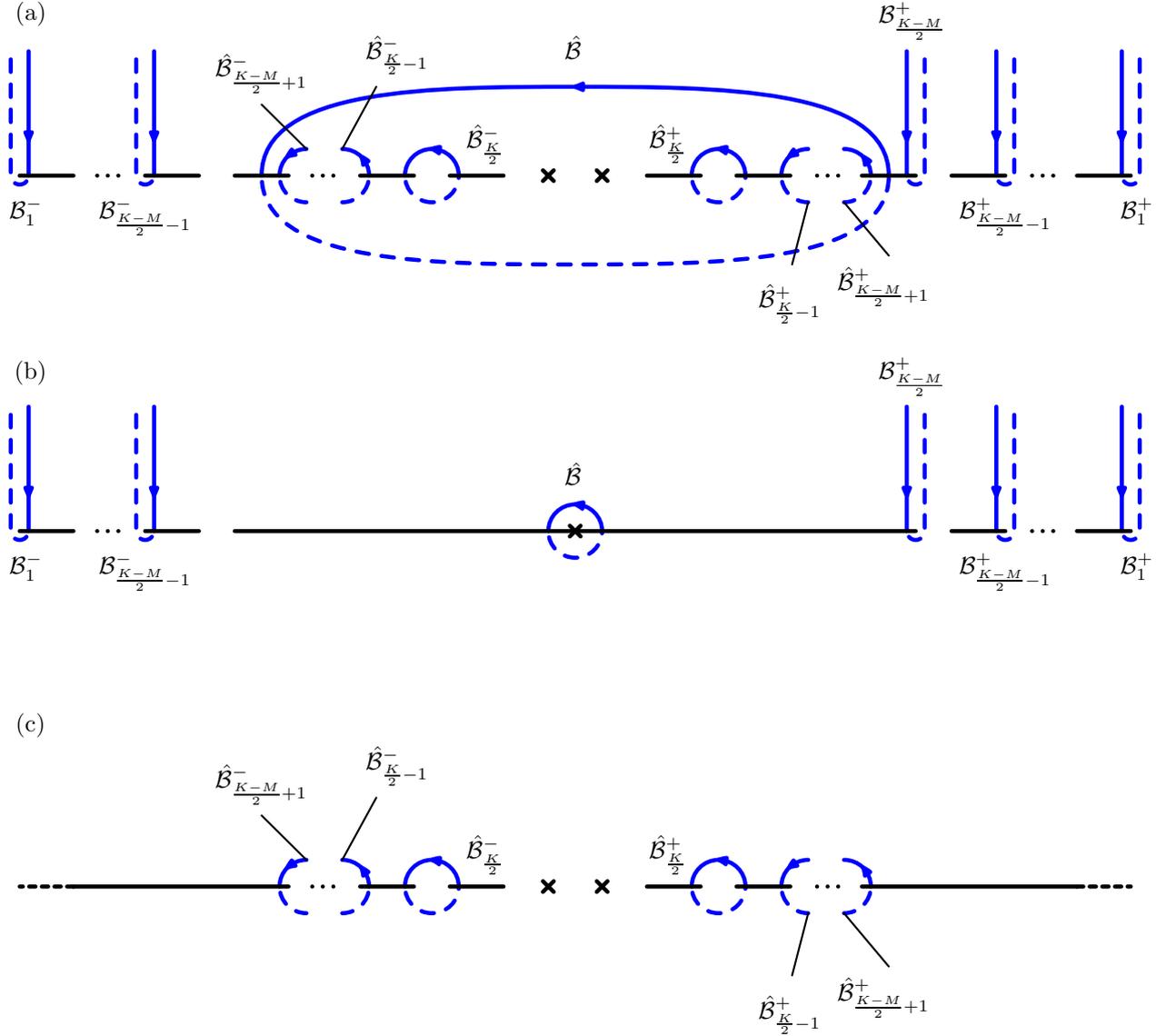}%
\caption{(a) The rearranged B-cycle configuration on $\Sigma$. Dashed blue lines indicate paths on the lower sheet throughout the picture. (b) The B-cycles on $\tilde{\Sigma}_1$; the marked point indicates the simple pole at $x=0$. (c) The B-cycles on $\tilde{\Sigma}_2$; the marked points indicate the double poles at $x = \pm 1$. The black dashed lines extend to infinity along the real axis.}
\label{fig:Bcycles}%
\end{figure}

First of all, it is convenient to rearrange the cycles on $\Sigma$ before we take the large-$S$ limit. The new equivalent configurations of the A-cycles and of the B-cycles are shown in Fig. \ref{fig:Acycles}(a) and \ref{fig:Bcycles}(a) respectively. In particular, we define $\hat{\mathcal{A}}_I^\pm = \sum_{J=I}^{K/2} \mathcal{A}_J^\pm$, for $I = (K-M)/2, \ldots, K/2$ and $\hat{\mathcal{A}}_0 = \sum_{J=(K-M)/2}^{K/2}(\mathcal{A}_J^+ + \mathcal{A}_J^-)$, together with $\hat{\mathcal{B}} = \mathcal{B}_{(K-M)/2}^- - \mathcal{B}_{(K-M)/2}^+$, $\hat{\mathcal{B}}_I^- = \mathcal{B}_{I-1}^- - \mathcal{B}_I^-$ and $\hat{\mathcal{B}}_I^+ = \mathcal{B}_I^+ - \mathcal{B}_{I-1}^+$, for $I = (K-M)/2 +1, \ldots, K/2$.

Basically, we have modified in a suitable way only the cycles associated with those cuts which will move to $\tilde{\Sigma}_2$ in the limit $\rho \to \infty$. A consistent set of period conditions on $\Sigma$ is then:
\begin{eqnarray}
  \oint_{\mathcal{A}_I^\pm} dp = 0 & \qquad & \oint_{\mathcal{B}_I^\pm} dp = 2 \pi n_I^\pm = \pm 2 \pi I 
   \qquad \textrm{for } I = 1, \ldots, \frac{K-M}{2} - 1 \nn \\
  \oint_{\tilde{\mathcal{A}}_0} dp = 0 & \qquad & \oint_{\hat{\mathcal{B}}} dp = - 2\pi (K-M) \nn \\
  \oint_{\hat{\mathcal{A}}_\frac{K-M}{2}^+} dp = 0 & \qquad & \oint_{\mathcal{B}_\frac{K-M}{2}^+} dp = \pi (K-M) \nn \\
  \oint_{\hat{\mathcal{A}}_J^\pm} dp = 0 & \qquad & \oint_{\hat{\mathcal{B}}_J^\pm} dp = 2 \pi \qquad \textrm{for } I = \frac{K-M}{2} +1, \ldots, K/2
\label{eq:new_period_conditions_S}
\end{eqnarray}
Fig. \ref{fig:Acycles}(b) and \ref{fig:Bcycles}(b) show which cycles survive on $\tilde{\Sigma}_1$ (identified as the region close to the branch points $a^{(j)}_\pm$) when we take $\rho$ to infinity, while Fig. \ref{fig:Acycles}(c) and \ref{fig:Bcycles}(c) show the configuration for $\tilde{\Sigma}_2$ (corresponding to the region close to the $b^{(j)}_\pm$) in the same limit.

As in the previous case, the spectral problem reduces to two (almost) separate problems on $\tilde{\Sigma}_1$ and $\tilde{\Sigma}_2$. On the first surface
\begin{equation}
 \tilde{\Sigma}_1 \:\: : \qquad \tilde{y}_1^2 = \prod_{k=1}^{K-M-1} (\tilde{x} - \tilde{a}^{(k)}_+) (\tilde{x} - \tilde{a}^{(k)}_-)
\label{eq:y1tilde_Mn0}
\end{equation}
we have the following limiting form of the differential
\begin{equation}
 d \tilde{p}_1 = - \frac{d \tilde{x}}{\tilde{y}_1} \sum_{l=M}^{K-2} \tilde{C}_l \tilde{x}^{l-M-1}
\label{eq:def_dp1tilde_Mn0}
\end{equation}
where $\tilde{x} = x / \rho$ as usual and we have rescaled the parameters $C_l$ as follows:
\begin{equation}
 C_l = \left\{
  \begin{array}{ll}
	 \tilde{C}_l \, \rho^{K-l-1} \:, & \textrm{for } l \geq M \\
	 \tilde{C}_l \, \rho^{K-M-1} \:, & \textrm{for } l < M
	\end{array}
	 \right.
\label{eq:rescaling_C_l_Mn0}
\end{equation}
This ensures that $d \tilde{p}_1$ has a simple pole at $\tilde{x} = 0$ and that none of these parameters disappears from both $d \tilde{p}_1$ and $d \tilde{p}_2$ due to suppression by negative powers of $\rho$. As we can see from Figures \ref{fig:Acycles}(b) and \ref{fig:Bcycles}(b), the differential is subject to the following period conditions, which are inherited from $\Sigma$ \eqref{eq:new_period_conditions_S}:
\begin{eqnarray}
  \oint_{\mathcal{A}_I^\pm} d \tilde{p}_1 = 0 & \qquad & \oint_{\mathcal{B}_I^\pm} d \tilde{p}_1 = 2 \pi n_I^\pm = \pm 2 \pi I 
   \qquad \textrm{for } I = 1, \ldots, \frac{K-M}{2} - 1 \nn \\
  \oint_{\tilde{\mathcal{A}}_0} d \tilde{p}_1 = 0 & \qquad & \oint_{\hat{\mathcal{B}}} d \tilde{p}_1 = - 2\pi (K-M) \nn \\
  & & \oint_{\mathcal{B}_\frac{K-M}{2}^+} d \tilde{p}_1 = \pi (K-M)
\label{eq:period_conditions_S1}
\end{eqnarray}
On $\tilde{\Sigma}_2$, which is parametrised as
\begin{equation}
 \tilde{\Sigma}_2 \:\: : \qquad \tilde{y}_2^2 = (x^2 - b^2) \prod_{i=1}^M (x - b^{(i)}_+) (x - b^{(i)}_-) \:,
\label{eq:eq:Sigma_2_tilde_Mn0}
\end{equation}
we find instead
\begin{equation}
 d \tilde{p}_2 = - \frac{dx}{\tilde{Q} \tilde{y}_2} \sum_{l=0}^M \tilde{C}_l x^l - \frac{\pi J}{\sqrt{\lambda}} \left[ 
  \frac{\tilde{y}_2 (1)}{(x-1)^2} + \frac{\tilde{y}_2 (-1)}{(x+1)^2} + 
   \frac{\tilde{y}'_2 (1)}{x-1} + \frac{\tilde{y}'_2 (-1)}{x+1} \right] \frac{dx}{\tilde{y}_2}
\label{eq:def_dp2tilde_Mn0}
\end{equation}
with $\tilde{Q}^2 = \tilde{y}_1^2(0)$. The corresponding period conditions are given by:
\begin{eqnarray}
  \oint_{\hat{\mathcal{A}}_J^\pm} d \tilde{p}_2 = 0 & \qquad & \oint_{\hat{\mathcal{B}}_J^\pm} d \tilde{p}_2 = 2 \pi \qquad \textrm{for } I = \frac{K-M}{2} +1, \ldots, K/2
\label{eq:period_conditions_S2}
\end{eqnarray}
as shown in Figures \ref{fig:Acycles}(c) and \ref{fig:Bcycles}(c). Only one period condition from the original set \eqref{eq:new_period_conditions_S} remains, namely:
\begin{equation}
 \oint_{\hat{\mathcal{A}}_\frac{K-M}{2}^+} dp = 0
\label{eq:match_cond}
\end{equation}
which involves the only cycle that survives both on $\tilde{\Sigma}_1$ and on $\tilde{\Sigma}_2$. This constraint then provides a relationship between the two problems associated with these surfaces and we will therefore refer to it as the ``matching condition''.

\subsubsection*{Explicit solution on $\tilde{\Sigma}_1$}

The period conditions on the first surface can be solved explicitly exactly as explained in \cite{DS}, the crucial point being once again that $d \tilde{p}_1$ only has a simple pole and no other singularities (apart from the branch points). We find:
\begin{equation}
 d \tilde{p}_1 = -i \frac{d \tilde{x}}{\tilde{x}^2} \frac{\mathbb{P}'_{K-M} \left( \frac{1}{\tilde{x}} \right)}{\sqrt{\mathbb{P}_{K-M}^2
  \left( \frac{1}{\tilde{x}} \right) -4}}
\label{eq:explicit_dp1tilde_Mn0}
\end{equation}
where
\begin{equation}
 \mathbb{P}_{K-M} \left( \frac{1}{\tilde{x}} \right) = 2 + \frac{\tilde{q}_2}{\tilde{x}^2} + \frac{\tilde{q}_3}{\tilde{x}^3}
  + \ldots + \frac{\tilde{q}_{K-M}}{\tilde{x}^{K-M}}
\label{eq:most_general_form_of_P_Mn0}
\end{equation}
and the parameters $\tilde{q}_j$ are related to the $\tilde{C}_l$ through
\begin{equation}
 \tilde{C}_l = - \frac{(K-l) \tilde{q}_{K-l}}{2 \sqrt{- \tilde{q}_2}}
\label{eq:Ctilde(qtilde)_Mn0}
\end{equation}
for $l = M, \ldots, K-2$.

As in the $M = 0$ case, the spectral curve coincides with the curve of a classical $SL(2,\mathbb{R})$ spin chain, whose length is now $K-M$, and $d \tilde{p}_1$ is the differential of the corresponding quasi-momentum. The $K-M-1$ parameters $\tilde{q}_j$ are the moduli associated with the first surface after the factorisation and they also represent the conserved charges of the spin chain.

With reference to Figure \ref{Gfig1b}, $\tilde{\Sigma}_1$ is the surface on the left-hand side of the bottom picture. Its main features are the $K-M-1$ branch cuts and the two singular contact points with $\tilde{\Sigma}_2$, which are located at $\tilde{x}=0^\pm$ (where $0^+$ lies on the top sheet, while $0^-$ lies on the bottom sheet).

\subsubsection*{Explicit solution on $\tilde{\Sigma}_2$}

The procedure on this surface is a bit more involved than it was in the previous case.

By imposing the matching condition \eqref{eq:match_cond} (which will be discussed in appendix \ref{sec:matching_condition}) at leading order, it is possible to show that, as in the $M=0$ case, the two innermost branch points on $\tilde{\Sigma}_2$ have to collide with the neighbouring double poles, i.e. that $b \to 1$ so that\footnote{The result \eqref{eq:def_dp2tilde_Mn0} still holds even though we now have a diverging factor $1/ \sqrt{1 - b^2}$ coming from $\tilde{y}'_2 (\pm 1)$, since corrections to that limit are suppressed by inverse powers of $\rho$ and hence a logarithmic divergence is too weak to make them $O(\rho^0)$.}:
\begin{equation}
 \frac{1}{\sqrt{1 - b^2}} \sim i \log \epsilon
\label{eq:sqrt(1-b^2)_lead_ord_Mn0}
\end{equation}
where $\epsilon = 1/\rho$. 

\begin{figure}
\centering
\psfrag{a}{\footnotesize{(a)}}
\psfrag{b}{\footnotesize{(b)}}
\psfrag{a1}{\footnotesize{$\hat{\mathcal{A}}_{\frac{K}{2}}^+$}}
\psfrag{a2}{\footnotesize{$\hat{\mathcal{A}}_{\frac{K}{2}-1}^+$}}
\psfrag{a3}{\footnotesize{$\hat{\mathcal{A}}_{\frac{K-M}{2}+1}^+$}}
\psfrag{a4}{\footnotesize{$\hat{\mathcal{A}}_{\frac{K-M}{2}}^+$}}
\psfrag{a5}{\footnotesize{$\hat{\mathcal{A}}_{\frac{K}{2}}^-$}}
\psfrag{a6}{\footnotesize{$\hat{\mathcal{A}}_{\frac{K}{2}-1}^-$}}
\psfrag{a7}{\footnotesize{$\hat{\mathcal{A}}_{\frac{K-M}{2}+1}^-$}}
\psfrag{c1}{\footnotesize{$c^{\left( 1 \right)}_+$}}
\psfrag{c2}{\footnotesize{$c^{\left( 2 \right)}_+$}}
\psfrag{c3}{\footnotesize{$c^{\left( \frac{M}{2} \right)}_+$}}
\psfrag{c4}{\footnotesize{$c^{\left( 1 \right)}_-$}}
\psfrag{c5}{\footnotesize{$c^{\left( 2 \right)}_-$}}
\psfrag{c6}{\footnotesize{$c^{\left( \frac{M}{2} \right)}_-$}}
\includegraphics[width=\columnwidth]{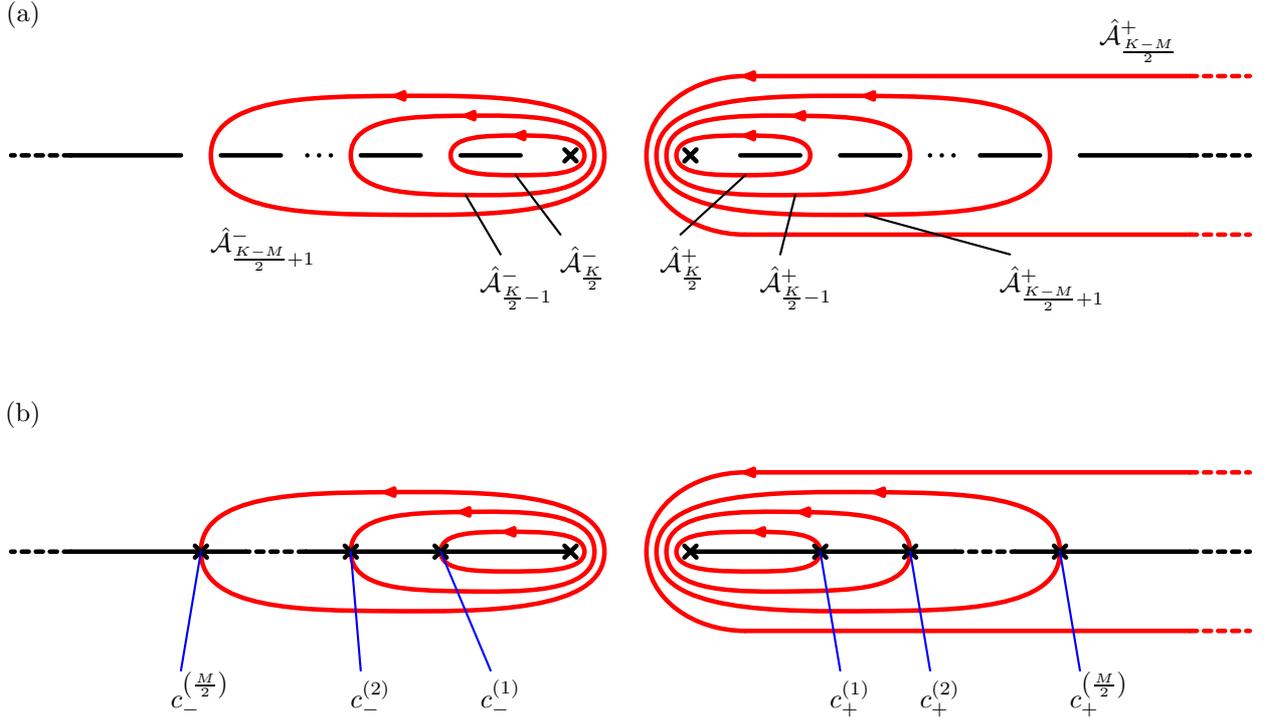}%
\caption{(a) The A-cycles on $\tilde{\Sigma}_2$ with the double poles inside them. (b) As $\rho \to \infty$, the A-cycles become pinched at the points $c^{(j)}_\pm$, for $j = 1, \ldots, M/2$.}
\label{fig:S2pinch}%
\end{figure}

We now focus our attention on the A-cycle conditions on $\tilde{\Sigma}_2$. First of all, we notice that, due to the fact that the double poles at $x = \pm 1$ have vanishing residues, $b \to 1$ does not imply that the contours are pinched at these singularities. In fact, even before the limit $\rho \to \infty$ is taken, we are free to rearrange the cycles so that they cross the real axis along the interval $-1 < x < 1$, where clearly there can be no pinching (see Fig. \ref{fig:S2pinch}(a)). It then follows that, even in the $\rho \to \infty$ limit, the contours $\hat{\mathcal{A}}_I^\pm$ do not touch any of the singularities of the differential. Therefore, the only diverging contribution the integrals receive comes from the factor $1/ \sqrt{1 - b^2}$ inside the integrand. For them to vanish, a second infinite contribution must arise in order to compensate.

This can only happen if all the branch points on $\tilde{\Sigma}_2$ coalesce in pairs as $\rho \to \infty$:
\begin{eqnarray}
 b^{(1)}_\pm , b^{(2)}_\pm & \to & c^{(1)}_\pm \nonumber\\
             & \vdots & \nonumber\\
 b^{(2j-1)}_\pm , b^{(2j)}_\pm & \to & c^{(j)}_\pm \nonumber\\
             & \vdots & \nonumber\\
 b^{(M-1)}_\pm , b^{(M)}_\pm & \to & c^{ \left( \frac{M}{2} \right) }_\pm
\label{eq:bps_coalesce}
\end{eqnarray}
The differential then develops a simple pole at each collision site $c^{(j)}_\pm$ and all the A-cycles are pinched at one (and only one) of these poles, as shown in Fig. \ref{fig:S2pinch}(b). Correspondingly, the genus of $\tilde{\Sigma}_2$ reduces from $M$ to $1$:
\begin{eqnarray}
 \tilde{y}_2 (x) & \to & \sqrt{x^2 - b^2} \prod_{j=1}^\frac{M}{2} ( x - c^{(j)}_+ ) ( x - c^{(j)}_- )  \qquad \textrm{as } \rho \to \infty
  \nonumber\\
                 & \equiv & \sqrt{x^2 - b^2} \: \hat{y}_2 (x)
\label{eq:simplification_of_y2tilde_after_collisions}
\end{eqnarray}
We can visualise the effect of this process with the help of Figure \ref{Gfig1b}. As the two surfaces separate from each other, the ``handles'' on $\tilde{\Sigma}_2$ (which lies on the right-hand side) collapse and each of them is replaced by a simple pole, represented as a blue dot in the picture.

This qualitative reasoning is already sufficient to determine the explicit form of $d \tilde{p}_2$. In particular, if we take into account the behaviour of all the branch points on this surface, we can write:
\begin{equation}
 d \tilde{p}_2 = \frac{h(x)}{\sqrt{x^2 - b^2}} dx + \ldots \qquad \textrm{as } \rho \to \infty
\label{eq:remove_sqrt(x^2-b^2)_from_dp2tilde_by_introducing_h(x)}
\end{equation}
where the dots denote terms which vanish in the limit considered\footnote{Strictly speaking, this is only true if corrections to \eqref{eq:bps_coalesce} are $O (\epsilon^\alpha)$ for some $\alpha > 0$, so that the logarithmically diverging factor $1/ \sqrt{1 - b^2}$ cannot generate $O(1)$ terms. It is possible to use the final explicit form of $d \tilde{p}_2$ in order to retrospectively check that this is the case.} and
\begin{equation}
 h(x) = - \frac{1}{\hat{y}_2 (x)} \left[ \frac{1}{\tilde{Q}} \sum_{l=0}^M \tilde{C}_l x^l + \frac{\pi J}{\sqrt{\lambda}} \frac{1}{\sqrt{1-b^2}}
  \left( \frac{\hat{y}_2 (1)}{x-1} - \frac{\hat{y}_2 (-1)}{x+1} \right) \right]
\label{eq:def_h(x)_for_limit_of_dp2tilde}
\end{equation}
is an analytic function which has simple poles at $x = \pm 1$ and $x = c^{(j)}_\pm$, for $j = 1, \ldots, M/2$. The limit of $h(x)$ as $x \to \infty$ and its residues at $x = \pm 1$ can be computed directly, while the residues at $x = c^{(j)}_\pm$ are determined by the B-period conditions in equation \eqref{eq:period_conditions_S2}. $h(x)$ can then be determined by analyticity constraints, yielding an explicit form for the differential of the quasi-momentum:
%\begin{multline}
% d \tilde{p}_2 = - \frac{2 \pi}{\sqrt{\lambda}} \frac{J}{\sqrt{1-b^2}} \frac{dx}{(x^2-1)^\frac{3}{2}} \\
%  + \left[  - \frac{K-M}{i} + \frac{1}{i} \sum_{j=1}^\frac{M}{2} \left( \frac{\sqrt{(c^{(j)}_+)^2 - 1}}{x-c^{(j)}_+} 
%   + \frac{\sqrt{(c^{(j)}_-)^2 - 1}}{x - c^{(j)}_-} \right) \right] \frac{dx}{\sqrt{x^2-1}}
%\label{eq:explicit_dp2tilde_no_match_cond}
%\end{multline}
\begin{equation}
 d \tilde{p}_2 = dw_0 + d \hat{w} + \sum_{j=1}^\frac{M}{2} (dw_j^+ + dw_j^-)
\label{eq:explicit_dp2tilde_no_match_cond}
\end{equation}
where we have defined
\begin{eqnarray}
 dw_0 & = & - \frac{2 \pi}{\sqrt{\lambda}} \frac{J}{\sqrt{1-b^2}} \frac{dx}{(x^2-1)^\frac{3}{2}} \nonumber\\
 dw_j^\pm & = & \frac{1}{i} \frac{\sqrt{(c^{(j)}_\pm)^2 - 1}}{x-c^{(j)}_\pm} \frac{dx}{\sqrt{x^2-1}} \nonumber\\
 d \hat{w} & = & - \frac{K-M}{i} \frac{dx}{\sqrt{x^2-1}}
\label{eq:def_dw0_dwhat_dwj^pm}
\end{eqnarray}
%We may now observe that the moduli on $\tilde{\Sigma}_2$ are given by the $M$ positions of the poles, $c^{(j)}_\pm$, for $j = 1, \ldots, M/2$ (since these are the only free parameters left in the final form of $d \tilde{p}_2$; $b$ is eliminated by the matching condition).
Finally, we can use this result in order to impose the matching condition up to $O(\rho^0)$ (see appendix \ref{sec:matching_condition}), obtaining:
\begin{equation}
 \frac{2 \pi}{\sqrt{\lambda}} \frac{J}{\sqrt{1-b^2}} = - i (K-M) \log \rho - i \log (\tilde{q}_{K-M}) - \frac{1}{2i} \sum_{j=1}^\frac{M}{2} \left[ T(c^{(j)}_+) + T(c^{(j)}_-) \right] - R
\label{eq:match_cond_without_infinite_sums}
\end{equation}
where $R$ is an undetermined moduli-independent constant and
\begin{equation}
 T(c) = \log \left( \frac{c - \sqrt{c^2-1}}{c + \sqrt{c^2-1}} \right)
\label{eq:def_T(c)}
\end{equation}
Now that $b$ has been eliminated by the matching condition, we observe that the only free parameters left in the final form of $d \tilde{p}_2$ are the $M$ positions of the poles, $c^{(j)}_\pm$, for $j = 1, \ldots, M/2$, which thus represent the  moduli on $\tilde{\Sigma}_2$. Adding these to the $K-M-1$ moduli from $\tilde{\Sigma}_1$, we obtain a $(K-1)$-dimensional moduli space of solutions.

As we can see in Figure \ref{Gfig1b}, the final configuration of
$\tilde{\Sigma}_2$ after the factorisation is characterised by two
singular contact points with $\tilde{\Sigma}_1$, located at $x =
\infty^\pm$, and $M$ simple poles on each sheet (represented as blue
dots). As the branch points approach $x = \pm 1$ the four double poles
above these points collide in pairs and the resulting differential $dw_{0}$ 
exhibits two double poles which coincide with branch points at $x = \pm 1$.

\subsubsection*{Conserved charges and filling fractions}

If we apply the asymptotic relations \eqref{asymp1} and \eqref{asymp2} to \eqref{eq:explicit_dp1tilde_Mn0} and \eqref{eq:explicit_dp2tilde_no_match_cond} respectively\footnote{We recall that the point $x=\infty$ on $\Sigma$ lies on $\tilde{\Sigma}_1$ after the factorisation, while $x=0$ on $\Sigma$ migrates to $\tilde{\Sigma}_2$.}, we obtain
\begin{equation}
 \Delta + S \simeq \frac{\sqrt{\lambda}}{2 \pi} \sqrt{- \tilde{q}_2} \,\, \rho
\label{eq:Delta+S_Mn0}
\end{equation}
and
%\begin{eqnarray}
% \Delta - S & = & \frac{2 \pi}{\sqrt{\lambda}} \left[ (K-M) \log \rho + \log (\tilde{q}_{K-M}) 
%  + \frac{1}{2} \sum_{j=1}^\frac{M}{2} \left( G( c^{(j)}_+ ) + G( c^{(j)}_- ) \right) \right. \nonumber\\
% & & \left. + \textrm{const.} \right]\label{eq:Delta-S_from_matching_v2_final}
%\end{eqnarray}
\begin{equation}
 \Delta - S \simeq \frac{\sqrt{\lambda}}{2 \pi} \left[ (K-M) \log \rho + \log (\tilde{q}_{K-M}) 
  + \sum_{j=1}^\frac{M}{2} \left( G( c^{(j)}_+ ) + G( c^{(j)}_- ) \right) + \textrm{const.} \right]
\label{eq:Delta-S_Mn0}
\end{equation}
where the constant is moduli-independent and we have introduced
\begin{equation}
 G(c) \equiv \frac{1}{2} \log \left( \frac{c + \sqrt{c^2-1}}{c - \sqrt{c^2-1}} \right) - \frac{\sqrt{c^2-1}}{c}
\label{eq:def_G(c)}
\end{equation}
Following the usual reasoning, we notice that $\Delta + S$ diverges faster than $\Delta - S$ and hence
\begin{equation}
 \Delta \simeq S \simeq \frac{\sqrt{\lambda}}{4 \pi} \sqrt{- \tilde{q}_2} \,\, \rho
\label{eq:S_as_rho_to_infty}
\end{equation}
which then implies
\begin{equation}
 \Delta - S \simeq \frac{\sqrt{\lambda}}{2 \pi} \left[ (K-M) \log S + \log \left( \frac{\tilde{q}_{K-M}}{(- \tilde{q}_2)^{K-M}} \right) 
  + \sum_{j=1}^\frac{M}{2} \left( G( c^{(j)}_+ ) + G( c^{(j)}_- ) \right) + \textrm{const.} \right]
\label{eq:Delta-S_Mn0_final}
\end{equation}
where the constant is again independent of the moduli.

Thus, we see that each simple pole on the surface $\tilde{\Sigma}_2$ is associated with an excitation yielding an $O(\sqrt{\lambda})$ and $O(S^0)$ contribution to the anomalous dimension. We propose that such excitations correspond to solitonic objects propagating along the finite-gap string solution with a worldsheet velocity $v = 1/c$, representing ``small'' spikes moving along the ``large'' spikes, as shown in Figure \ref{SGMfig7}. Because of the original ordering of the branch points, we have $|c^{(j)}_\pm| > 1$, $\forall j$, and thus $-1 < v < 1$. The conserved energy of an excitation is then
\begin{equation}
 E(v) = \frac{\sqrt{\lambda}}{2 \pi} \left[ \frac{1}{2} \log \left( \frac{1 + \sqrt{1-v^2}}{1 - \sqrt{1-v^2}} \right) - \sqrt{1-v^2} \right]
\label{eq:E(v)}
\end{equation}

\paragraph{}

In order to complete the picture, we need to implement the semiclassical quantisation conditions, which require the filling fractions, defined in equation \eqref{eq:filling_fractions}, to be integers.

On $\tilde{\Sigma}_1$, we impose
\begin{eqnarray}
 - \frac{1}{2 \pi i} \frac{1}{\sqrt{- \tilde{q}_2}} \oint_{\mathcal{A}_I^\pm} \tilde{x} \, d \tilde{p}_1 & = & \frac{l_I^\pm}{S}
  \qquad \textrm{for } I = 1, \ldots, \frac{K-M}{2} - 1 \nonumber\\
 - \frac{1}{2 \pi i} \frac{1}{\sqrt{- \tilde{q}_2}} \oint_{\tilde{\mathcal{A}}_0} \tilde{x} \, d \tilde{p}_1 & = & \frac{l_0}{S}
\label{eq:filling_fractions_S1}
\end{eqnarray}
with $l_I^\pm, l_0 \in \mathbb{Z}$, which leads to the usual quantisation of the spectrum associated with spikes approaching the boundary (i.e. ``large'' spikes), as discussed in \cite{DS}. In particular, the moduli $\tilde{q}_j$ become discretised: $\tilde{q}_j = \tilde{q}_j (l_I^\pm, l_0)$.

On $\tilde{\Sigma}_2$, we redefine the remaining filling fractions by replacing the contours $\mathcal{A}_I^\pm$ with $\hat{\mathcal{A}_I^\pm}$, for $I = (K-M)/2 + 1, \ldots, K/2$. We may compute the relevant contour integrals at leading order by using the explicit expression for $d \tilde{p}_2$ \eqref{eq:explicit_dp2tilde_no_match_cond}\footnote{This step requires particular care: the contours are pinched at the poles $c^{(j)}_\pm$, and hence the filling fractions must be regulated. For this purpose, before we take $\rho \to \infty$, we convert the integral along $\hat{\mathcal{A}}_{K/2 - (j - 1)}^\pm$ into an open chain starting at $b^{(2j-1)}_\pm$ on one side of the cut, intersecting the real axis between the double poles, and ending at $b^{(2j-1)}_\pm$ on the other side of the cut, for $j = 1, \ldots, M/2$. We then write $b^{(2j-1)}_\pm = c^{(j)}_\pm \mp \eta^{(j)}_\pm$, according to \eqref{eq:bps_coalesce}, and use \eqref{eq:explicit_dp2tilde_no_match_cond} to compute the integral. Finally, the behaviour of the regulator $\eta^{(j)}_\pm$ can be determined by imposing, at leading order, the A-period condition involving the same contour $\hat{\mathcal{A}}_{K/2 - (j - 1)}^\pm$, again turning the latter into an open chain and using \eqref{eq:explicit_dp2tilde_no_match_cond}.}:
\begin{equation}
 \mathcal{S}_j^\pm = - \frac{\sqrt{\lambda}}{4 \pi} \frac{1}{2 \pi i} \oint_{\hat{\mathcal{A}}^\pm_{\frac{K}{2}-(j-1)}} \left( x + \frac{1}{x}
  \right) dp = \mathcal{S} (c^{(j)}_\pm)
\label{eq:filling_fractions_S2}
\end{equation}
where
\begin{equation}
 \mathcal{S} (c) = \frac{\sqrt{\lambda}}{4 \pi^2} \left[ \sqrt{c^2-1} + \tan^{-1} \left( \frac{1}{\sqrt{c^2-1}} \right) \right] (K-M) \log \rho
\label{eq:def_S(c)}
\end{equation}
We now wish to interpret the constraint $\mathcal{S} (c) \in \mathbb{Z}$ as the Bohr-Sommerfeld quantisation condition for a particle of momentum $P(c)$ in a box of length $L \simeq (K-M) \log \rho$ (which is equal, at leading order, to the length of the finite-gap string solution\footnote{The infinite GKP string has length $L = (\Delta - S) (2 \pi / \sqrt{\lambda}) \sim \log S$. As shown in \cite{DL1,GH}, if we bend this solution into an arc of the Kruczenski string or if we add a small spike propagating along it, we only introduce $O(1)$ corrections to this relation. Hence, it is reasonable to expect that $L \simeq (\Delta - S) (2 \pi / \sqrt{\lambda})$ at leading order even for a general solution consisting of several arcs and small spikes such as the one shown in Fig. \ref{SGMfig7}. We may then extract the length of the finite-gap solution from the spectrum \eqref{eq:Delta-S_Mn0_final}.}). Such a condition would read $P(c) L / (2 \pi) \in \mathbb{Z}$, leading us to the identification
\begin{equation}
 \frac{P(c) L}{2 \pi} = \mathcal{S} (c)
\label{eq:identify_P(c)_and_S(c)}
\end{equation}
This allows us to introduce a conserved momentum associated with each excitation, which we now express as a function of the velocity $v$:
\begin{equation}
 P(v) = \frac{\sqrt{\lambda}}{2 \pi} \left[ \frac{\sqrt{1-v^2}}{v} - \mathrm{Tan}^{-1} \left( \frac{\sqrt{1-v^2}}{v} \right) \right]
\label{eq:P(v)}
\end{equation}
where $\mathrm{Tan}^{-1}$ is the principal branch of the inverse tangent and we have chosen the origin of the scale of momenta so that $P(\pm 1) = 0$, which is justified since $E(\pm 1) = 0$ and hence the excitations disappear for these extremal values of the velocity.

We now observe that the dispersion relation of the excitations, given by equations \eqref{eq:E(v)} and \eqref{eq:P(v)}, is the same as the dispersion relation of the ``Giant Holes'' \eqref{disp2}.

As a final remark, the case we have considered here is that of equal numbers of branch cuts moving to $\tilde{\Sigma}_2$ from the two halves of the real axis. We have restricted ourselves to this case for simplicity, but clearly the above reasoning still applies when different numbers of cuts remain close to the origin on the two sides as $\rho \to \infty$, and even when $K$ is odd. Hence, the finite-gap construction can describe solutions with arbitrary numbers of ``small'' and ``large'' spikes.

\section{Conclusions}

We have studied the spectral curves corresponding to a general class
of $K$-gap string solutions living on $AdS_3 \times S^1$. 
Our analysis shows that these curves factorise into two different
surfaces as the $AdS_3$ angular momentum $S$ diverges. The first
surface, $\tilde{\Sigma}_1$ has arbitrary genus $K-M-2$, 
which controls the leading diverging behaviour of the spectrum:
\begin{equation}
 \Delta - S \simeq \frac{\sqrt{\lambda}}{2 \pi} \left[ (K-M) \log S + 
\log \left( \frac{\tilde{q}_{K-M}}{(- \tilde{q}_2)^{K-M}} \right) \right] + O(1) \nn
\end{equation}
According to the hypothesis of \cite{DS}, we expect the corresponding
string solutions to develop $K-M$ ``large'' spikes which approach the
boundary of anti-de Sitter space as $S \to \infty$. The moduli on
$\tilde{\Sigma}_1$ govern the dynamics of these objects. This part of
the spectrum precisely matches the contribution of $(K-M)$ ``large''
holes in the dual $\mathrm{SL} (2, \mathbb{R})$ 
spin chain.

The second surface $\tilde{\Sigma}_2$ always has genus zero and
contains an arbitrary number $M$ of simple poles of the differential
of the quasi-momentum. Each of these extra poles introduces a finite
$O(\sqrt{\lambda},S^0)$ contribution $E(v)$ to the anomalous dimension
$\Delta - S$, where the parameter $v$ is determined by the position of
the pole. A conserved momentum $P(v)$ can also be associated with the
pole through the corresponding filling fraction and the dispersion
relation $E(P)$ is the same as was observed for the ``Giant Holes'' in
\cite{GH}, which in turn coincides with the dispersion relation for
the ``small'' holes of the gauge theory spin chain in the limit of high momentum $P \gg 1$.

We find that each pole on the second surface corresponds to a
solitonic excitation propagating along the string with worldsheet
velocity $v$, carrying conserved energy $E(v)$ and momentum
$P(v)$. This excitation should correspond to a ``small'' spike that
does not extend towards the boundary as $S \to \infty$ and whose
worldsheet position coincides with the position of a soliton of the
complex sinh-Gordon equation, appearing in the Pohlmeyer-reduced
picture of classical string theory in $AdS_3 \times S^1$. In the limit
$S\rightarrow \infty$, the dynamics
of these excitations is determined by the moduli on $\tilde{\Sigma}_2$
and is completely decoupled 
from the dynamics of ``large'' spikes.
The spectrum
\begin{equation}
 \Delta - S \simeq \frac{\sqrt{\lambda}}{2 \pi} \left[ (K-M) \log S + \log \left( \frac{\tilde{q}_{K-M}}{(- \tilde{q}_2)^{K-M}} \right) \right]
  + \sum_{j=1}^M E(v_j) + \textrm{const.} \nn
\end{equation}
provides a semiclassical description of the string dual of an
$\mathrm{SL} (2, \mathbb{R})$ spin chain configuration with $M$
``small'' 
holes and $K-M$ ``large'' holes. The corresponding string solution is
therefore expected to consist of $M$ small solitonic spikes propagating on the
background of $K-M$ large spikes which approach the boundary (see
Figure 2). It would be interesting to find an explicit interpolation
between the semiclassical string spectrum discussed here and the
corresponding gauge theory spin chain using the exact Asymptotic Bethe
Ansatz Equations of \cite{BS, BES}.

\appendix

\section{The matching condition}
\label{sec:matching_condition}

In this section we sketch how to deal with the only period condition which involves both surfaces $\tilde{\Sigma}_1$ and $\tilde{\Sigma}_2$ at the same time:
\begin{equation}
\oint_{\hat{\mathcal{A}}_\frac{K-M}{2}^+} dp = 0
\label{eq:repeat_match_cond}
\end{equation}
For our initial analysis, we only need the final expression for $d \tilde{p}_1$ \eqref{eq:explicit_dp1tilde_Mn0}. We start by splitting the integral into two separate contributions:
\begin{equation}
 \oint_{\hat{\mathcal{A}}_{(K-M)/2}^+} dp = 2 I_1 + 2 I_2 = 0
\label{eq:splitting_integral_for_matching_condition_Mn0}
\end{equation}
with:
\begin{eqnarray}
 I_1 & \equiv & - \frac{1}{2} \int_{b^-}^{b^+} \frac{dx}{y} \sum_{l=M}^{K-2} C_l x^l \nonumber\\
 I_2 & \equiv & - \frac{1}{2} \int_{(a^{(1)}_+)^+}^{(a^{(1)}_+)^-} \frac{dx}{y} \left[ \sum_{l=0}^{M-1} C_l x^l + \frac{\pi J}{\sqrt{\lambda}}
        \right. \nonumber\\
     &        & \left. \phantom{\sum_{l=M}^{K-2}} \times \left( \frac{y_+}{(x-1)^2} + \frac{y_-}{(x+1)^2} + \frac{y'_+}{x-1} + \frac{y'_-}{x+1}
                 \right) \right]
\label{eq:def_I1_I2_Mn0}
\end{eqnarray}
where we have opened up the contour at the left endpoint for $I_1$ and at the right endpoint for $I_2$, turning the integrals into open chains which start at $b$ and respectively $a^{(1)}_+$ on one side of the corresponding cut and end at $b$ and $a^{(1)}_+$ on the other side.

Both integrals can be evaluated on $\tilde{\Sigma}_1$, by introducing the change of variables $x = \rho \tilde{x}$. This will introduce factors of the type $\sqrt{\tilde{x} - \epsilon b^{(j)}_\pm}$ into $y$, which can be treated by making use of the binomial expansion\footnote{Strictly speaking, these expansions only converge for $\tilde{x} > \textrm{max } \{ \epsilon b^{(M)}_+, - \epsilon b^{(M)}_- \}$. One way around the problem is to introduce the series, at first restricting ourselves to the region of convergence. After this, we swap the sum with the integral and only then we remove the regulator by letting $\tilde{x}$ reach the endpoint of integration, $\tilde{x} = \epsilon b$. This is what is done below. Another way is to divide the problematic region of the contour into several segments, each joining two consecutive points of the set $\{ \epsilon b^{(j)}_+, - \epsilon b^{(j)}_- \}$. We can then introduce appropriate converging binomial expansions (either of the form \eqref{eq:binomial_expansion_for_sqrt_in_y_on_S1} or with $k_j^\pm$ and $- 1/2 - k_j^\pm$ interchanged) for each interval. The final result is the same.}:
\begin{equation}
 \frac{1}{\sqrt{\tilde{x} - \epsilon b^{(j)}_\pm}} = \sum_{k_j^\pm = 0}^\infty \binom{- \frac{1}{2}}{k_j^\pm} (- \epsilon b^{(j)}_\pm )^{k_j^\pm}
  \tilde{x}^{- \frac{1}{2} - {k_j^\pm}}
\label{eq:binomial_expansion_for_sqrt_in_y_on_S1}
\end{equation}
One can also similarly expand the terms $(x \pm 1)^{-1}$ and $(x \pm 1)^{-2}$ appearing in $I_2$.

In the case of $I_1$, we have the integral of an infinite sum over $2(M+1)$ indices (one for each square root factor we had to expand), which we can write as:
\begin{eqnarray}
 I_1 & = & - \frac{1}{2} \sum_{k_0^\pm, \ldots, k_M^\pm = 0}^\infty \epsilon^{k_\mathrm{tot}} \left[ \prod_{j=0}^M \binom{- \frac{1}{2}}{k_j^+}
  \binom{- \frac{1}{2}}{k_j^-} (- b^{(j)}_+ )^{k_j^+} (- b^{(j)}_- )^{k_j^-} \right] \nonumber\\
     &   & \quad \times \int_{(\epsilon b)^-}^{(\epsilon b)^+} \frac{d \tilde{x}}{\tilde{y}_1} 
            \sum_{l=M}^{K-2} \tilde{C}_l \tilde{x}^{l -(M+1) - k_\mathrm{tot}}
\label{eq:I1_after_expanding_sqrts_Mn0}
\end{eqnarray}
with $k_\mathrm{tot} = \sum_{j=0}^M (k_j^+ + k_j^-)$ and $b^{(0)}_\pm = \pm b$. The remaining integral can now be calculated straightforwardly. For $k_\mathrm{tot} > 0$, the leading order is easily seen to be $O(1)$ ($\tilde{y}_1 (\tilde{x}) \simeq \tilde{y}_1 (0) = \tilde{Q}$). We indicate the sum of all these constant terms as $\hat{I}_1 (c^{(j)}_\pm)$ (at leading order, all the $b^{(j)}_\pm$ reduce to one of the $c^{(j)}_\pm$, according to \eqref{eq:bps_coalesce}, while $b = 1$, thus this expression only depends on the moduli $c^{(j)}_\pm$).

The remaining $k_\mathrm{tot} = 0$ term can be rewritten as:
\begin{equation}
 - \frac{1}{2} \int_{(\epsilon b)^-}^{(\epsilon b)^+} \frac{d \tilde{x}}{\tilde{y}_1} 
            \sum_{l=M}^{K-2} \tilde{C}_l \tilde{x}^{l -M-1}
 = \frac{1}{2} \int_{(\epsilon b)^-}^{(\epsilon b)^+} d \tilde{p}_1
 = \frac{1}{2} [ 2 \tilde{p}_1 (\epsilon b) + 2 \pi (K-M) ]
\label{eq:writing_ktot=0_term_in_I1_in_terms_of_p1tilde_Mn0}
\end{equation}
where we have used the discontinuity properties of $\tilde{p}_1$ in the last step. \eqref{eq:explicit_dp1tilde_Mn0} then implies:
\begin{equation}
 I_1 = - i (K-M) \log \epsilon + i \log \left( \frac{\tilde{q}_{K-M}}{b^{K-M}} \right) + \pi (K-M) + \hat{I}_1 (c^{(j)}_\pm) + O(\epsilon)
\label{eq:final_form_of_I1_Mn0}
\end{equation}
For the moment, the only feature of $I_1$ we are interested in is the fact that it diverges as $\log \epsilon$ in the limit $\epsilon \to 0$. Due to the A-cycle condition \eqref{eq:splitting_integral_for_matching_condition_Mn0}, $I_2$ must then also diverge in the same limit. However, a similar analysis to the one carried out for $I_1$ shows that $I_2$\footnote{This analysis requires us to split the contour for $I_2$ at $b$, not at $a^{(1)}_+$ as indicated above. This is also a legitimate operation. The above definition will be useful when analysing $I_2$ on $\tilde{\Sigma}_2$.} is $O(1) + O( \epsilon )$. The solution lies in the fact that some of the $O(1)$ terms are proportional to $1/ \sqrt{1 - b^2}$ (these terms originate from $y'(\pm 1)$). $I_2 \sim i \log \epsilon$ then implies:
\begin{equation}
 \frac{1}{\sqrt{1 - b^2}} \sim i \log \epsilon
\label{eq:asymptotic_behaviour_of_sqrt(1-b^2)_from_matching_condition_Mn0}
\end{equation}
which is the result we referred to in equation \eqref{eq:sqrt(1-b^2)_lead_ord_Mn0}. As we saw in the following discussion, this implies that the branch points on $\tilde{\Sigma}_2$ must coalesce and that $d \tilde{p}_2$ must take the form \eqref{eq:explicit_dp2tilde_no_match_cond}.

We can then proceed to evaluate $I_2$ on $\tilde{\Sigma}_2$, i.e. without changing variables to $\tilde{x}$. The steps are similar to what we did in the other coordinates. In particular, we now have to expand the following factors coming from $y$\footnote{Note that ensuring the convergence of the binomial series again requires particular care. We need $|x \epsilon / \tilde{a}^{(j)}_\pm| < 1$, for $j = 1, \ldots, K-M-1$. If we shrink the contour onto the real axis, we can safely assume $|x| \in (0, \rho \tilde{a}^{(1)}_+]$ over the whole domain of integration. The problem arises near the upper limit: $\tilde{a}^{(1)}_+ / |\tilde{a}^{(j)}_\pm|$ is not necessarily less than $1$ for all $j$. However, $\mathrm{min} \{ |\tilde{a}^{(1)}_-| , \tilde{a}^{(1)}_+ \} / |\tilde{a}^{(j)}_\pm| \leq 1$, $\forall j$, and hence we should actually evaluate the matching condition on $\hat{\mathcal{A}}^-_{(K-M)/2}$ instead of $\hat{\mathcal{A}}^+_{(K-M)/2}$ when $|\tilde{a}^{(1)}_-| < \tilde{a}^{(1)}_+$. Nonetheless, all the calculations would work in exactly the same way and the final version \eqref{eq:match_cond_without_infinite_sums} of the matching condition would be identical (as we may guess from the fact that it doesn't depend on $\tilde{a}^{(1)}_+$). Alternatively, we could apply the same reasoning as we discussed for \eqref{eq:binomial_expansion_for_sqrt_in_y_on_S1}.}:
\begin{equation}
 \frac{1}{\sqrt{x \epsilon - \tilde{a}^{(j)}_\pm}} = \sum_{k_j^\pm = 0}^\infty \binom{- \frac{1}{2}}{k_j^\pm}
  (- \tilde{a}^{(j)}_\pm )^{- \frac{1}{2} - k_j^\pm} (x \epsilon)^{k_j^\pm}
\label{eq:binomial_expansion_for_sqrt_in_y_S2}
\end{equation}
Again the integral turns into an infinite sum of integrals. This time, however, the $k_\mathrm{tot} > 0$ part is $O( \epsilon \log \epsilon)$ and hence it does not contribute. We are left with the $k_\mathrm{tot} = 0$ term only:
\begin{eqnarray}
 I_2 & \simeq & - \frac{1}{2} \int_{(\rho \tilde{a}^{(1)}_+)^+}^{(\rho \tilde{a}^{(1)}_+)^-} \frac{dx}{\tilde{y}_2} \left[ 
            \frac{1}{\tilde{Q}} \sum_{l=0}^{M-1} \tilde{C}_l x^l \right. \nonumber\\
     &   & \left. \phantom{\sum_{l=0}^{M-1}} + \frac{\pi J}{\sqrt{\lambda}} \left( \frac{\tilde{y}_2 (1)}{(x-1)^2}
            + \frac{\tilde{y}_2 (-1)}{(x+1)^2} + \frac{\tilde{y}'_2 (1)}{x-1} + \frac{\tilde{y}'_2 (-1)}{x+1} \right) \right]
\label{eq:I2_after_eliminating_sqrts}
\end{eqnarray}
We now observe that the differential appearing in the above integral is \emph{not} the part of $dp$ which contributes to $d \tilde{p}_2$ as $\rho \to \infty$. In particular, it is missing the $l = M$ term in the first sum. We will call the limit as $\rho \to \infty$ of this ``incomplete'' differential $d \hat{p}_2$.

In order to determine $d \hat{p}_2$, one may go through the same steps which led us to $d \tilde{p}_2$, most of which are identical to what we have already seen. The only difference is that now:
\begin{equation}
 \lim_{x \to \infty} h(x) = 0
\label{eq:limit_as_x_to_inf_of_h(x)_for_dp2hat}
\end{equation}
which has the effect of killing the only term in \eqref{eq:explicit_dp2tilde_no_match_cond} that is not regular at infinity:
\begin{equation}
 d \hat{p}_2 = dw_0 + \sum_{j=1}^\frac{M}{2} (dw_j^+ + dw_j^-)
\label{eq:explicit_dp2hat}
\end{equation}
Hence, we may write:
\begin{equation}
 I_2 = \frac{1}{2} \int_{(\rho \tilde{a}^{(1)}_+)^+}^{(\rho \tilde{a}^{(1)}_+)^-} [ d \hat{p}_2 (x) + O(\epsilon \log \epsilon) ]
  \qquad \textrm{as } \epsilon = 1/ \rho \to 0
\label{eq:I_2_in_terms_of_dp2hat}
\end{equation}
where now $O( \epsilon \log \epsilon)$, as a function of $x$, is manifestly integrable on the whole domain of integration, even as $\epsilon \to 0$, due to the fact that both the starting differential appearing in \eqref{eq:I2_after_eliminating_sqrts} and $d \hat{p}_2$ are integrable. This means that its primitive does not diverge as $x \to \infty$ and hence this contribution still vanishes even after integration. Therefore we may neglect it when computing $I_2$ up to $O(1)$:
\begin{eqnarray}
 I_2 & = & \frac{1}{2} \int_{(\rho \tilde{a}^{(1)}_+)^+}^{(\rho \tilde{a}^{(1)}_+)^-} d \hat{p}_2 (x) + \ldots \nonumber\\
     & = & \frac{2 \pi}{\sqrt{\lambda}} \frac{J}{\sqrt{1-b^2}} + \frac{1}{2i} \sum_{j=1}^\frac{M}{2} 
            \left[ T( c^{(j)}_+ ) + T( c^{(j)}_- ) \right] + \frac{\hat{L}_2}{2} + \ldots
\label{eq:I2_up_to_O(1)}
\end{eqnarray}
where $T(c)$ was defined in \eqref{eq:def_T(c)}, the dots denote corrections which vanish in the limit $\rho \to \infty$ and $\hat{L}_2$ is the constant part of the discontinuity of $\hat{p}_2$ across its cut in the region $x > c^{(M/2)}_+$ ($\hat{p}_2 (x + i \epsilon) + \hat{p}_2 (x - i \epsilon) = \hat{L}_2$ for $x \in \mathbb{R}$, $x > c^{(M/2)}_+$).

Thus, the matching condition, up to $O(1)$ in $\epsilon$, yields:
\begin{multline}
 \frac{2 \pi}{\sqrt{\lambda}} \frac{J}{\sqrt{1-b^2}} = i (K-M) \log \epsilon - i \log \left( \frac{\tilde{q}_{K-M}}{b^{K-M}} \right) 
  - \pi (K-M) - \hat{I}_1 (c^{(j)}_\pm) \\
 - \frac{1}{2i} \sum_{j=1}^\frac{M}{2} \left[ T( c^{(j)}_+ ) + T( c^{(j)}_- ) \right] - \frac{\hat{L}_2}{2}
\label{eq:matching_condition_up_to_O(1)_S1}
\end{multline}
Before substituting back into $d \tilde{p}_2$, we will impose the matching condition through a slightly different procedure, which will yield an alternative version of the above expression. By comparing the two versions, we will then see that the expression simplifies.

The idea is now to evaluate \eqref{eq:repeat_match_cond} entirely on $\tilde{\Sigma}_2$, without splitting the integral into two contributions. The first step is to open up the contour at $a^{(1)}_+$:
\begin{equation}
 \int^{(a^{(1)}_+)^+}_{(a^{(1)}_+)^-} dp = 0
\label{eq:opening_up_contour_match_cond_S2}
\end{equation}
where the points ${(a^{(1)}_+)^\pm}$ lie at $x = a^{(1)}_+$ on the top ($+$) and bottom ($-$) sheet.

Again, the integrand develops factors of the type \eqref{eq:binomial_expansion_for_sqrt_in_y_S2} and we may turn the integral into an infinite sum of integrals by using the same binomial expansion. The $k_\mathrm{tot} > 0$ term is $O(1)$, as well as the part of the $k_\mathrm{tot} = 0$ contribution which depends on the $\tilde{C}_l$, for $l = M+1, \ldots, K-2$. We indicate the total of these two contributions as $2 \hat{I}_2 (\tilde{q}_j)$ ($\hat{I}_2$ depends on the $\tilde{a}^{(j)}_\pm$ and on the $\tilde{C}_l$, for $l = M, \ldots, K-2$; in the $S \to \infty$ limit, both sets of parameters only depend on the $\tilde{q}_j$ through \eqref{eq:explicit_dp1tilde_Mn0} and \eqref{eq:Ctilde(qtilde)_Mn0}).

The remaining integrand reduces to $d \tilde{p}_2$ as $\rho \to \infty$, so that \eqref{eq:opening_up_contour_match_cond_S2} may be written as:
\begin{equation}
 2 \hat{I}_2 (\tilde{q}_j) + \int^{(\rho \tilde{a}^{(1)}_+)^+}_{(\rho \tilde{a}^{(1)}_+)^-} d \tilde{p}_2 = 0
\label{eq:match_cond_with_dp2tilde_S2}
\end{equation}
and we may then use \eqref{eq:explicit_dp2tilde_no_match_cond} to recast the matching condition on $\tilde{\Sigma}_2$ into the following form:
\begin{equation}
 \frac{2 \pi}{\sqrt{\lambda}} \frac{J}{\sqrt{1-b^2}} = - \hat{I}_2 (\tilde{q}_j) + \frac{K-M}{i} \log (2 \rho \tilde{a}^{(1)}_+)
  - \frac{1}{2i} \sum_{j=1}^\frac{M}{2} \left[ T(c^{(j)}_+) + T(c^{(j)}_-) \right] - \frac{\tilde{L}_2}{2}
\label{eq:match_cond_on_S2}
\end{equation}
where the moduli-independent constant $\tilde{L}_2$ is obtained from the discontinuity of $\tilde{p}_2$ at its cut: $\tilde{p}_2 (x+i \epsilon) + \tilde{p}_2 (x - i \epsilon) = \tilde{L}_2$, for $x \in \mathbb{R}$, $x > c^{(M/2)}_+$.

By equating the RHS of \eqref{eq:matching_condition_up_to_O(1)_S1} and \eqref{eq:match_cond_on_S2}, we find:
\begin{multline}
 \hat{I}_2 (\tilde{q}_j) + i (K-M) \log (2 \tilde{a}^{(1)}_+) - i \log ( \tilde{q}_{K-M} ) = \\ \hat{I}_1 (c^{(j)}_\pm) - i (K-M) \log b + \pi (K-M) + \frac{\hat{L}_2}{2} - \frac{\tilde{L}_2}{2}
\label{eq:comparing_2_versions_of_match_cond}
\end{multline}
At this point, we observe that the $K-1$ independent degrees of freedom, which parametrise this class of finite-gap solutions after all the period conditions have been implemented, are given by $\tilde{q}_j$, $j = 2, \ldots, K-M$, and $c^{(k)}_\pm$, $k = 1, \ldots, M/2$. The quasi-momentum $\tilde{p}_1$ on $\tilde{\Sigma}_1$ is completely determined by the $\tilde{q}_j$, while $\tilde{p}_2$ on $\tilde{\Sigma}_2$ is completely determined by the $c^{(k)}_\pm$. The matching condition does not impose any extra constraint on these $K-1$ moduli; instead, it determines the behaviour of $b$ as $\rho \to \infty$, i.e. $b = 1 + \ldots$, where the dots represent vanishing corrections (i.e. it determines one of the parameters of the curve as a function of the moduli, so that in the end the only free parameters left are the moduli themselves).

If we look at equation \eqref{eq:comparing_2_versions_of_match_cond} in the limit $\rho \to \infty$, and we neglect the $\log b$ term, which vanishes at $O(1)$, we easily see that the LHS only depends on the $\tilde{q}_j$, while the RHS only depends on the $c^{(k)}_\pm$\footnote{All of this strictly holds only in the $S \to \infty$ limit.}. As we have just explained, this equation cannot be used in order to eliminate one of the moduli in terms of the others, and hence it can only be satisfied if both sides are equal to a moduli-independent constant, which we call $R'$.

This, together with any of the expressions \eqref{eq:matching_condition_up_to_O(1)_S1} and \eqref{eq:match_cond_on_S2}, yields:
\begin{equation}
 \frac{2 \pi}{\sqrt{\lambda}} \frac{J}{\sqrt{1-b^2}} = - i (K-M) \log \rho - i \log (\tilde{q}_{K-M}) - \frac{1}{2i} \sum_{j=1}^\frac{M}{2} \left[ T(c^{(j)}_+) + T(c^{(j)}_-) \right] - R' - \frac{\tilde{L}_2}{2}
\label{eq:match_cond_final_R'}
\end{equation}
which is easily seen to reduce to \eqref{eq:match_cond_without_infinite_sums} under the identification $R = R' + \tilde{L}_2 / 2$.


\begin{thebibliography}{99}


\bibitem{MZ}
  J.~A.~Minahan and K.~Zarembo,
  ``The Bethe-ansatz for N = 4 super Yang-Mills,''
  JHEP {\bf 0303} (2003) 013
   [arXiv:hep-th/0212208].

\bibitem{B}   
 N.~Beisert and M.~Staudacher,
  ``The N = 4 SYM integrable super spin chain,''
  Nucl.\ Phys.\ B {\bf 670} (2003) 439
  [arXiv:hep-th/0307042]. \\
N.~Beisert, C.~Kristjansen and M.~Staudacher,
  ``The dilatation operator of N = 4 super Yang-Mills theory,''
  Nucl.\ Phys.\ B {\bf 664} (2003) 131
  [arXiv:hep-th/0303060]. 

\bibitem{MSW}
  G.~Mandal, N.~V.~Suryanarayana and S.~R.~Wadia,
  %``Aspects of semiclassical strings in AdS(5),''
  Phys.\ Lett.\  B {\bf 543} (2002) 81
  [arXiv:hep-th/0206103].

\bibitem{BPR}
  I.~Bena, J.~Polchinski and R.~Roiban,
  ``Hidden symmetries of the AdS(5) x S**5 superstring,''
  Phys.\ Rev.\ D {\bf 69}, 046002 (2004)

\bibitem{Serban} D.~Serban,
  ``Integrability and the AdS/CFT correspondence,''
  arXiv:1003.4214 [Unknown].

%\bibitem{BMN} D.~E.~Berenstein, J.~M.~Maldacena and H.~S.~Nastase,
%  ``Strings in flat space and pp waves from N = 4 super Yang Mills,''
%  JHEP {\bf 0204} (2002) 013
%  [arXiv:hep-th/0202021].

\bibitem{BDS}  N.~Beisert, V.~Dippel and M.~Staudacher,
  ``A novel long range spin chain and planar N = 4 super Yang-Mills,''
  JHEP {\bf 0407} (2004) 075
  [arXiv:hep-th/0405001].

%\bibitem{Staudacher}
%  M.~Staudacher,
%  ``The factorized S-matrix of CFT/AdS,''
%  JHEP {\bf 0505} (2005) 054
%  [arXiv:hep-th/0412188].
  %%CITATION = HEP-TH 0412188;%%

\bibitem{BS} N.~Beisert and M.~Staudacher,
  ``Long-range PSU(2,2$|$4) Bethe ansaetze for gauge theory and strings,''
  Nucl.\ Phys.\ B {\bf 727} (2005) 1
  [arXiv:hep-th/0504190].

\bibitem{B1} N.~Beisert, 
  ``The su(2|2) dynamic S-matrix,''
  Adv.\ Theor.\ Math.\ Phys.\  {\bf 12} (2008) 945
  [arXiv:hep-th/0511082].


\bibitem{HM}
  D.~M.~Hofman and J.~M.~Maldacena,
  ``Giant magnons,''
  J.\ Phys.\ A  {\bf 39}, 13095 (2006)
  [arXiv:hep-th/0604135].
  
\bibitem{AGV} L.~F.~Alday, D.~Gaiotto, J.~Maldacena, A.~Sever and P.~Vieira,
  ``An Operator Product Expansion for Polygonal null Wilson Loops,''
  arXiv:1006.2788 [hep-th].
  
\bibitem{GH}
  N.~Dorey and M.~Losi,
  ``Giant Holes,'' 
J.\ Phys.\ A  {\bf 43} (2010) 285402
  [arXiv:1001.4750 [hep-th]].
 

%\bibitem{GKV}
%  N.~Gromov, V.~Kazakov and P.~Vieira,
%  ``Exact Spectrum of Anomalous Dimensions of Planar N=4 Supersymmetric
%  Yang-Mills Theory,''
%  Phys.\ Rev.\ Lett.\  {\bf 103} (2009) 131601
%  arXiv:0901.3753 [hep-th]. \\ 
%N.~Gromov, V.~Kazakov, A.~Kozak and P.~Vieira,
%  ``Integrability for the Full Spectrum of Planar AdS/CFT II,''
%  arXiv:0902.4458 [hep-th]. \\  
%N.~Gromov, V.~Kazakov and P.~Vieira,
%  ``Exact AdS/CFT spectrum: Konishi dimension at any coupling,''
%  arXiv:0906.4240 [hep-th].

%\bibitem{BFT}
%  D.~Bombardelli, D.~Fioravanti and R.~Tateo,
%  ``Thermodynamic Bethe Ansatz for planar AdS/CFT: a proposal,''
%  J.\ Phys.\ A  {\bf 42} (2009) 375401
%  arXiv:0902.3930 [hep-th].

%\bibitem{AF1} G.~Arutyunov and S.~Frolov,
%  ``Thermodynamic Bethe Ansatz for the $AdS_5 x S^5$ Mirror Model,''
%  JHEP {\bf 0905} (2009) 068
%  arXiv:0903.0141 [hep-th]. \\
%  G.~Arutyunov and S.~Frolov,
%  ``Simplified TBA equations of the $AdS_5 x S^5$ mirror model,''
%  JHEP {\bf 0911} (2009) 019
%  arXiv:0907.2647 [hep-th].


\bibitem{BGK}
  A.~V.~Belitsky, A.~S.~Gorsky and G.~P.~Korchemsky,
  ``Logarithmic scaling in gauge / string correspondence,''
  Nucl.\ Phys.\  B {\bf 748}, 24 (2006)
  [arXiv:hep-th/0601112].

\bibitem{FS} L.~Freyhult, A.~Rej and M.~Staudacher,
  ``A Generalized Scaling Function for AdS/CFT,''
  J.\ Stat.\ Mech.\  {\bf 0807}, P07015 (2008)
  [arXiv:0712.2743 [hep-th]].


\bibitem{GKP}
  S.~S.~Gubser, I.~R.~Klebanov and A.~M.~Polyakov,
  ``A semi-classical limit of the gauge/string correspondence,''
  Nucl.\ Phys.\  B {\bf 636}, 99 (2002)
  [arXiv:hep-th/0204051].

\bibitem{BES}
  N.~Beisert, B.~Eden and M.~Staudacher,
  ``Transcendentality and crossing,''
  J.\ Stat.\ Mech.\  {\bf 0701} (2007) P021
  [arXiv:hep-th/0610251].


\bibitem{AM2}
  L.~F.~Alday and J.~M.~Maldacena,
 ``Comments on operators with large spin,''
  JHEP {\bf 0711} (2007) 019
  arXiv:0708.0672 [hep-th].

\bibitem{Kruc} M.~Kruczenski,
  ``Spiky strings and single trace operators in gauge theories,''
  JHEP {\bf 0508} (2005) 014
  [arXiv:hep-th/0410226].

\bibitem{DS}
  N.~Dorey,
  ``A Spin Chain from String Theory,''
  Acta Phys.\ Polon.\  B {\bf 39}, 3081 (2008)
  arXiv:0805.4387 [hep-th].

\bibitem{DL1} N.~Dorey and M.~Losi,
 ``Spiky Strings and Spin Chains,''
  arXiv:0812.1704 [hep-th].

\bibitem{Jev1}
  A.~Jevicki, K.~Jin, C.~Kalousios and A.~Volovich,
 % ``Generating AdS String Solutions,''
  JHEP {\bf 0803} (2008) 032
  arXiv:0712.1193 [hep-th]. 
  
\bibitem{Pohlmeyer}
  K.~Pohlmeyer,
  %``Integrable Hamiltonian Systems And Interactions Through Quadratic
  %Constraints,''
  Commun.\ Math.\ Phys.\  {\bf 46} (1976) 207.

\bibitem{otherPohl} M.~Grigoriev and A.~A.~Tseytlin,
  %``On reduced models for superstrings on AdS_n x S^n,''
  Int.\ J.\ Mod.\ Phys.\  A {\bf 23} (2008) 2107
  arXiv:0806.2623 [hep-th]. \\ 
  J.~L.~Miramontes,
  %``Pohlmeyer reduction revisited,''
  JHEP {\bf 0810} (2008) 087
  arXiv:0808.3365 [hep-th].

\bibitem{Jev2} 
  A.~Jevicki and K.~Jin,
  %``Solitons and AdS String Solutions,''
  Int.\ J.\ Mod.\ Phys.\  A {\bf 23}, 2289 (2008)
  arXiv:0804.0412 [hep-th]. \\
  A.~Jevicki and K.~Jin,
  %``Moduli Dynamics of AdS_3 Strings,''
  JHEP {\bf 0906}, 064 (2009)
  arXiv:0903.3389 [hep-th]. \\
  A.~Jevicki and K.~Jin,
  %``AdS String: Classical Solutions and Moduli Dynamics,''
  arXiv:1001.5301 [hep-th].



%\bibitem{AMs}  L.~F.~Alday and J.~M.~Maldacena,
%  ``Gluon scattering amplitudes at strong coupling,''
%  JHEP {\bf 0706}, 064 (2007)
%  arXiv:0705.0303 [hep-th]. \\
%L.~F.~Alday and J.~Maldacena,
%  ``Minimal surfaces in AdS and the eight-gluon scattering amplitude 
%  at strong
%  coupling,''
%  arXiv:0903.4707 [hep-th]. \\
%  L.~F.~Alday and J.~Maldacena,
%  ``Null polygonal Wilson loops and minimal surfaces in 
%  Anti-de-Sitter space,''
%  JHEP {\bf 0911}, 082 (2009)
%  arXiv:0904.0663 [hep-th].





\bibitem{KMMZ}
  V.~A.~Kazakov, A.~Marshakov, J.~A.~Minahan and K.~Zarembo,
  ``Classical / quantum integrability in AdS/CFT,''
  JHEP {\bf 0405} (2004) 024
  [arXiv:hep-th/0402207].

\bibitem{KZ} V.~A.~Kazakov and K.~Zarembo,
 ``Classical / quantum integrability in non-compact sector of AdS/CFT,''
  JHEP {\bf 0410} (2004) 060
  [arXiv:hep-th/0410105].


\bibitem{DV1}
  N.~Dorey and B.~Vicedo,
  ``On the dynamics of finite-gap solutions in classical string theory,''
  JHEP {\bf 0607} (2006) 014
  [arXiv:hep-th/0601194].

\bibitem{DV2} 
  N.~Dorey and B.~Vicedo,
  ``A symplectic structure for string theory on integrable backgrounds,''
  JHEP {\bf 0703} (2007) 045
  [arXiv:hep-th/0606287].

\end{thebibliography}
\end{document}